\documentclass[aps,letterpaper,prb,twocolumn,notitlepage,showpacs,superscriptaddress,nobalancelastpage
    amsmath]{revtex4-1}
\usepackage{bbm,color,amsmath,amssymb,graphicx}
\usepackage[pdftex,
            pdfauthor={Yang-Le Wu},
            pdftitle={Matrix Product State Representation of Non-Abelian Quasiholes}]{hyperref}

\begin{document}

\title{Matrix product state representation of non-Abelian quasiholes}

\author{Yang-Le Wu}
\affiliation{Condensed Matter Theory Center and Joint Quantum Institute,
Department of Physics, University of Maryland, College Park, Maryland 20742, USA}
\author{B. Estienne}
\affiliation{Sorbonne Universit\'es, UPMC Univ Paris 06, UMR 7589, LPTHE, F-75005, Paris, France}
\affiliation{CNRS, UMR 7589, LPTHE, F-75005, Paris, France}
\author{N. Regnault}
\affiliation{Department of Physics, Princeton University, Princeton, New Jersey 08544, USA}
\affiliation{Laboratoire Pierre Aigrain, Ecole Normale Sup\'erieure-PSL 
Research University, CNRS, Universit\'e Pierre et Marie Curie-Sorbonne 
Universit\'es, Universit\'e Paris Diderot-Sorbonne Paris Cit\'e, 24 rue 
Lhomond, 75231 Paris Cedex 05, France}
\author{B. Andrei Bernevig}
\affiliation{Department of Physics, Princeton University, Princeton, New Jersey 08544, USA}

\date{\today}

\begin{abstract}
We provide a detailed explanation of the formalism necessary to construct matrix product states for non-Abelian quasiholes 
in fractional quantum Hall model states.
Our construction yields an efficient representation of the wave functions with 
conformal-block normalization and monodromy, and complements the matrix product state representation of fractional quantum Hall ground states.
\end{abstract}
\maketitle

\section{Introduction}

The fractional quantum Hall (FQH) effect~\cite{Tsui82:FQH}
is the archetypal system for the emergence of 
topological order~\cite{Wen95:TopoOrder} in condensed matter physics.
Due to the presence of strong correlations, the theoretical understanding of its 
microscopic properties heavily relies on finite-size 
numerics.~\cite{Laughlin83:Nobel,Haldane83:Sphere,
Tserkovnyak03:MR,Prodan09:Braiding,Baraban09:MR}
Such calculations are limited to rather small system sizes, as the
many-body Hilbert space grows exponentially with the system area.
An ingenious way to circumvent this problem comes from the recent 
development~\cite{Dubail12:MPS,Zaletel12:MPS,Estienne13:MPS,Estienne13:MPSLong,
Wu14:Quasihole,Estienne14:Gaffnian,Zaletel13:DMRG}
of exact matrix product states~\cite{Fannes92:MPS,Schollwock11:DMRG-MPS}
(MPS) for a large class of FQH model wave functions derived from conformal 
field theory (CFT) correlators~\cite{Belavin84:BPZ,Fubini91:FQH-CFT,Moore91:MR}.
Exploiting the area law of quantum entanglement,~\cite{Eisert10:AreaLaw}
the MPS factorization enables efficient calculation of physical observables 
and gives access to much larger system sizes than previously attainable. 

In a recent paper,~\cite{Wu14:Quasihole} we applied this technique to study 
quasihole excitations in the Moore-Read,~\cite{Moore91:MR} the 
Gaffnian,~\cite{Simon07:Gaffnian} and the $\mathbb{Z}_3$ 
Read-Rezayi~\cite{Read99:RR} states.
These exotic quasiparticles were conjectured to be non-Abelian anyons, and
they constitute the most striking manifestation of the topological order in 
the FQH liquids.~\cite{Kitaev03:TQC,Nayak08:RMP}
A hallmark of their non-Abelian character lies in the topological degeneracy
of multiquasihole states. 
For each set of fixed quasihole positions, there exists a quasidegenerate 
subspace of states, protected against local perturbations,
and braiding the quasiholes induces unitary transformations
over this degenerate subspace.
These transformations depend only on the topology of the braids, rather than 
their actual shapes, and those induced by distinct braids do not commute.
In Ref.~\onlinecite{Wu14:Quasihole}, using the MPS representation, we 
explicitly demonstrated for the first time the Fibonacci nature of the 
$\mathbb{Z}_3$ Read-Rezayi quasiholes from a microscopic calculation.
We estimated the quasihole radii and quantified the length scales
associated with the exponential convergence of the braiding statistics,
but did not provide details regarding the construction of the 
quasihole wavefunctions.
In this paper, we discuss in detail the novel technical aspects
of the MPS representation for the non-Abelian quasiholes.

The construction of the exact MPS is based on the rewriting of FQH model 
wave functions as conformal correlators.~\cite{Fubini91:FQH-CFT,Moore91:MR}
This elegant formalism provides a particularly nice way to resolve the 
topological degeneracy of non-Abelian quasiholes~\cite{Moore91:MR} in terms of 
the so-called conformal blocks.~\cite{Moore89:CFT,DiFrancesco99:Yellow}
Each conformal-block wave function is indexed not only by the quasihole 
positions, but also by a tree of topological charge labels specifying the different
fusion channels of the quasiholes.
Enumerating all the possible fusion tree labelings compatible with a given theory
generates a complete basis over the degenerate subspace.
A special benefit of the conformal-block basis is the explicit manifestation 
of the \emph{putative} braiding statistics in the analytic structure of the 
wave functions.~\cite{Nayak96:SO2n,Ardonne07:RR}
Specifically, as a function of the complex quasihole coordinates, the 
conformal blocks display branch-cut singularities emanating from the quasihole 
centers.
The monodromy matrix associated with crossing the branch cuts is 
conjectured~\cite{Moore91:MR}
to coincide with the corresponding quasihole braiding matrix,
up to an Abelian Aharonov-Bohm phase due to the magnetic field.
This conjecture rests on the observation that the overlaps between conformal 
blocks resemble the partition function of a classical 
plasma~\cite{Laughlin83:Nobel} with pinned 
quasihole charges, a peculiar feature of the quasihole-dependent 
\emph{normalization} of the conformal 
blocks.~\cite{Arovas84:Statistics}
This link eliminates the need to directly integrate the non-Abelian Berry 
connection to compute the braiding 
matrix.~\cite{Gurarie97:Plasma,Read09:Adiabatic,Bonderson11:Plasma}
To demonstrate that the braiding statistics is indeed captured by the 
monodromy of conformal blocks, we only have to establish (through a 
microscopic calculation) that the plasma is in a screening phase.
This simplification led to the analytic identification (with the assumption of plasma screening) of the 
Moore-Read quasiholes as Ising anyons in Ref.~\onlinecite{Bonderson11:Plasma},
and also played a crucial role in our numerical demonstration~\cite{Wu14:Quasihole}
of the $\mathbb{Z}_3$ Read-Rezayi quasiholes as Fibonacci anyons.
The main purpose of the current paper is to explain how to translate the 
conformal blocks into calculation-friendly MPS form while preserving
the two highly desirable features, 
namely, the monodromy structure and the plasma normalization.

The paper is organized as follows.
In Sec.~\ref{sec:review}, we provide a pedagogical review on the construction 
of quantum Hall MPS in the absence of non-Abelian quasiholes, and in 
particular, we derive the plasma normalization from conformal correlators on 
the cylinder.
And as a precursor, we also discuss the MPS representation of Abelian 
quasiholes.
In Sec.~\ref{sec:non-Abelian} we proceed to the non-Abelian case.
We explain step by step how to construct MPS for the conformal-block wave 
functions with quasiholes in the bulk.
A large part of the discussion is devoted to the derivation of the subtle commutation 
rules between a non-Abelian quasihole insertion and the electron operators.
We provide explicit recipes for the Moore-Read,~\cite{Moore91:MR}
the Gaffnian,~\cite{Simon07:Gaffnian}, and the 
$\mathbb{Z}_3$ Read-Rezayi~\cite{Read99:RR} states.
Technical details of the construction are addressed in the appendices.

\section{Matrix product states from conformal correlators}\label{sec:review}

Before discussing the non-Abelian quasiholes, we first review the
construction~\cite{Dubail12:MPS,Zaletel12:MPS,Estienne13:MPS,Estienne13:MPSLong}
of the quantum Hall matrix product states (MPS) from conformal correlators.
In preparation for the later discussion of the quasihole wave functions,
we pay special attention to preserving the normalization of the conformal 
correlator.

We consider model wave functions in the lowest Landau level
constructed from chiral conformal correlators.~\cite{Fubini91:FQH-CFT,Moore91:MR}
In this formalism, an electron at position $z$ is represented by a primary 
field insertion $\mathcal{V}(z)$,
and the conformal correlator
\begin{equation}\label{eq:correlator-primitive}
\Big\langle \mathcal{V}(z_1)\mathcal{V}(z_2)\cdots \mathcal{V}(z_n)
\Big\rangle
\end{equation}
can be viewed as a many-body wave function $\Psi(z_1,z_2,\ldots,z_n)$.
Here and hereafter, the single brackets $\langle\,\cdot\,\rangle$ denote a CFT 
correlation function,
in contrast to the double brackets $\langle\!\langle\,\cdot\,\rangle\!\rangle$
representing the states (wave functions) of the physical electrons.
For example, the Laughlin wave function~\cite{Laughlin83:Nobel} at filling 
$\nu$ can be described~\cite{Fubini91:FQH-CFT} in terms of a
massless free boson $\phi$.
The electron operator $\mathcal{V}(z)$ is the normal-ordered exponential
\begin{equation}\label{eq:Ve-Laughlin}
\mathcal{V}(z)=\,:\!e^{i\frac{1}{\sqrt{\nu}}\phi(z)}\!:.
\end{equation}
Using the propagator 
${\langle\phi(z)\phi(z')\rangle=-\log(z-z')}$ in the plane, we 
recover from 
Eq.~\eqref{eq:correlator-primitive} the familiar Laughlin wave function 
${\prod_{i<j}(z_i-z_j)^{1/\nu}}$.
For more complicated quantum Hall states (see Sec.~\ref{sec:non-Abelian}), the 
corresponding CFT has a direct-product structure,~\cite{Moore91:MR}
where in addition to the free boson, we also have a separate so-called 
``neutral'' CFT.

From now on we adopt the cylinder geometry~\cite{Rezayi94:Cylinder} with 
finite perimeter $L_y$.
The complex coordinate ${z=x+iy}$ has $x$ running along the cylinder axis and $y$ 
around its perimeter.
For convenience, we set the magnetic length to unity,
and we define the inverse cylinder radius
\begin{equation}
\gamma=\frac{2\pi}{L_y}.
\end{equation}
The many-body wave function $\Psi(z_1,z_2,\ldots,z_n)$ is given by
the conformal correlator in Eq.~\eqref{eq:correlator-primitive} evaluated in 
the cylinder geometry, which can be mapped the usual planar 
geometry through the conformal transformation $z\rightarrow e^{\gamma z}$.

Interpreting the $x$ coordinate as the imaginary time, the CFT Hamiltonian
is given by $\gamma(\hat{L}_0-\frac{1}{24}c)$,
with $\hat{L}_0$ being the Virasoro generator for dilations
and $c$ being the chiral central charge.
For the direct-product theory of a neutral CFT and a free boson, we have the 
decomposition
\begin{equation}\label{eq:L0-total}
\hat{L}_0=\hat{L}_0^\text{neut}+\hat{L}_0^\text{boson}.
\end{equation}
In this section, we focus on the boson part.
The mode expansion of the chiral field $\phi$ on the cylinder is given by
\begin{equation}\label{eq:boson-mode-expansion}
\phi(z)=\hat{\phi}_0-i\gamma z\,\hat{a}_0
+i\sum_{n\neq 0}\frac{1}{n}\,\hat{a}_ne^{-n\gamma z}.
\end{equation}
Here, the $\hat{a}_n$ modes of the U(1) current satisfy the Heisenberg algebra
\begin{equation}
[\hat{a}_n,\hat{a}_m]=n\,\delta_{n+m,0},
\end{equation}
while $\hat{\phi}_0$ is the canonical conjugate to the zero mode $\hat{a}_0$,
\begin{equation}\label{eq:boson-zero-mode}
[\hat{\phi}_0,\hat{a}_0]= i.
\end{equation}
And the dilation operator for the free boson is given by
\begin{equation}\label{eq:boson-L0}
\hat{L}_0^\text{boson}=\sum_{m>0}\hat{a}_{-m}\hat{a}_m+\frac{1}{2}\hat{a}_0^2.
\end{equation}
The $\hat{a}_0$ operator measures the U(1) charge in unit of $\sqrt{\nu}$ 
times the electron charge, in the sense that
\begin{equation}
[\mathcal{V}(z)]^{-1}\,\hat{a}_0\,\mathcal{V}(z)=\hat{a}_0+\frac{1}{\sqrt{\nu}}.
\end{equation}
The zero mode operators $\hat{\phi}_0$ and $\hat{a}_0$ are
decoupled from (commute with) 
the ladder operators $\hat{a}_{n\neq 0}$.
As a result, the free boson Hilbert space can be split into sectors labeled by 
the U(1) charge $\hat{a}_0$, with $\hat{\phi}_0$ coupling different sectors.
The primary state 
$|Q\rangle={:\!e^{i Q\hat{\phi}_0}\!:}|\mathbbm{1}\rangle$
has U(1) charge $Q$, and the corresponding
Hilbert space sector with charge $Q$ is spanned by the descendants of 
$|Q\rangle$ under the boson modes $\{\hat{a}_{n<0}\}$.

\subsection{Background charge and gauge choice}

The MPS is a tensor factorization of the second-quantized amplitudes of the 
many-body wave function $\Psi(z_1,z_2,\ldots,z_N)$ in 
Eq.~\eqref{eq:correlator-primitive}.~\cite{Dubail12:MPS,Zaletel12:MPS}
The first step of the construction is to obtain $\Psi$
in the occupation-number basis.
We choose the Landau gauge for the magnetic field, and work with the
orbitals labeled by the wave number $j\in\mathbb{Z}$ in the $y$ direction:
\begin{equation}\label{eq:Landau-orbital}
\begin{aligned}
\psi_j(x,y)
&=\frac{1}{(\sqrt{\pi}L_y)^{\frac{1}{2}}}\,e^{i\gamma jy}e^{-\frac{1}{2}(x-\gamma j)^2}\\
&=\frac{e^{-\frac{1}{2}\gamma^2 j^2}}{(\sqrt{\pi}L_y)^{\frac{1}{2}}}\,
e^{\gamma jz}e^{-\frac{1}{2}x^2}.
\end{aligned}
\end{equation}
These one-body states take the form of a 
holomorphic function in $z$ times a Gaussian in $x$.
Due to the chirality of the electron operators $\mathcal{V}(z)$, the conformal 
correlator in Eq.~\eqref{eq:correlator-primitive} does not produce the
(non-holomorphic) Gaussian factor, and thus does not yet qualify as a many-body 
wave function in the lowest Landau level (in any gauge).
Fortunately, the Gaussian factor can be generated naturally
by spreading the neutralizing background 
charge for the boson field $\phi$ \emph{uniformly}~\cite{Moore91:MR} on the cylinder.
This amounts to inserting another (non-primary) field
\begin{equation}\label{eq:bc}
\mathcal{O}_\text{bc}=\,\,
:\!\exp\left(-i\frac{\sqrt{\nu}}{2\pi}\int\mathrm{d}^2w\,\,\phi(w)\right)\!:
\end{equation}
into the conformal correlator, representing the neutralizing background charge 
at filling $\nu$.
Here, the integration is performed over the cylinder surface, and the normal 
ordering removes unwanted interactions between background charges at different 
locations.
However, as discussed in Ref.~\onlinecite{Moore91:MR}, this extra insertion of the background charge
has a side effect: in addition to the desirable Gaussian factor, it also 
introduces a non-holomorphic gauge factor.
Taken altogether, the cylinder many-body wave function in the Landau gauge is 
given by~\cite{Zaletel12:MPS}
\begin{equation}\label{eq:correlator-cylinder}
\Psi(z_1,\ldots,z_n)=
e^{i\sum_i x_i y_i}
\Big\langle \mathcal{V}(z_1)\cdots \mathcal{V}(z_n)\,\,
\mathcal{O}_\text{bc}
\Big\rangle.
\end{equation}
This relation is proved in Appendix~\ref{sec:background-charge}.

\subsection{Occupation-number basis}\label{sec:occupation-number}
We now try to expand the above wave function in the occupation-number basis.
Since each Landau orbital is a momentum eigenstate, we can extract the 
second-quantized amplitudes through a Fourier transform in the $y$ direction.
Notice that the one-body wave function $\psi_j$ reduces to a simple plane wave 
along the orbital center $x=\gamma j$,
\begin{equation}
\psi_j(\gamma j,y)=
\frac{1}{(\sqrt{\pi}L_y)^{\frac{1}{2}}}
e^{i\gamma jy}.
\end{equation}
Taking advantage of this,~\cite{Zaletel12:MPS} we place the Fourier 
integration contours along the orbital centers, and express the amplitude
associated with the occupied orbitals $\{j_1,j_2,\cdots,j_n\}$ as
\begin{equation}\label{eq:correlator-psi-j}
\begin{aligned}
&\Psi_{j_1,j_2,\ldots,j_n}\\
&=
\prod_i
\int_0^{L_y}\frac{\mathrm{d}y_i}{L_y}\,
e^{-i\gamma j_i y_i}\,
\Psi(\gamma j_1+iy_1,\ldots,\gamma j_n+iy_n)\\
&=
\prod_i
\int_0^{L_y}\frac{\mathrm{d}y_i}{L_y}\,
\Big\langle \mathcal{V}(\gamma j_1+iy_1)\cdots 
\mathcal{V}(\gamma j_n+iy_n)\,\,
\mathcal{O}_\text{bc}
\Big\rangle,
\end{aligned}
\end{equation}
up to a constant normalization factor.
Here and hereafter we consider only the fermionic quantum Hall states.
Notice that at $x_i=\gamma j_i$, the gauge transformation $e^{i x_iy_i}$ in 
Eq.~\eqref{eq:correlator-cylinder} cancels the Fourier factor 
$e^{-i\gamma j_iy_i}$.

The next step is to rewrite the above expression in terms of the occupation 
numbers $m_j=0,1$ of each Landau orbital $j$.
We work in the operator formalism of the CFT,
and interpret the $x$ direction along the cylinder axis as the imaginary time
and the perpendicular $y$ direction as the space direction.
For simplicity, we first consider only the electron operators, and postpone 
the treatment of the background charge operator.
We are free to pick the index ordering of the occupied orbitals.
Choosing $j_1>j_2>\cdots>j_n$ ensures the time ordering of the electron operators 
and allows us to convert the correlator into an operator expression
\begin{multline}\label{eq:operator-formalism}
\Big\langle \mathcal{V}(\gamma j_1+iy_1)\cdots 
\mathcal{V}(\gamma j_n+iy_n)
\Big\rangle
=\\
\langle\mathrm{out}|
\hat{\mathcal{V}}(\gamma j_1+iy_1)\cdots 
\hat{\mathcal{V}}(\gamma j_n+iy_n)
|\mathrm{in}\rangle.
\end{multline}
We use over-hats on the right-hand side to highlight the operator nature of the 
insertions, and we can set the in- and the out-states to the vacuum.
Thanks to the conformal invariance,
the $x$ dependence of the primary field insertion $\mathcal{V}(x+iy)$ can be 
isolated,
\begin{equation}\label{eq:dilation}
\hat{\mathcal{V}}(x+iy)
=e^{x\gamma \hat{L}_0}\hat{\mathcal{V}}(iy)e^{-x\gamma\hat{L}_0},
\end{equation}
with $\gamma \hat{L}_0$ being the CFT Hamiltonian on the cylinder.
We define the zero-mode of the electron operator as
\begin{equation}\label{eq:V-zero-mode}
\hat{\mathcal{V}}_0=\int_{-L_y/2}^{L_y/2}\frac{\mathrm{d}y}{L_y}\,\hat{\mathcal{V}}(iy).
\end{equation}
Without worrying about the background charge for now, we can rewrite
Eq.~\eqref{eq:correlator-psi-j} as
\begin{multline}\label{eq:time-evolution-punctured}
\langle\mathrm{out}|
\,\hat{\mathcal{V}}_0\,
e^{-(j_1-j_2)\gamma^2\hat{L}_0}
\,\hat{\mathcal{V}}_0\,
e^{-(j_2-j_3)\gamma^2\hat{L}_0}
\cdots\\
\cdots e^{-(j_{n-1}-j_n)\gamma^2\hat{L}_0}
\,\hat{\mathcal{V}}_0\,
|\mathrm{in}\rangle,
\end{multline}
up to factors that depend only on the energy of the in- and the out-state
boundaries.
The above expression is an imaginary time evolution along the cylinder axis, 
punctured by the electron zero-mode operators at the center of each occupied 
orbital.~\cite{Zaletel12:MPS}
To assign the time evolution to individual orbitals,
we define
\begin{equation}\label{eq:U-naive}
\hat{U}(s)=e^{-s\gamma\hat{L}_0},
\end{equation}
which advances in imaginary time any CFT state by $s$ 
along the cylinder.
Finally, we can write the second-quantized amplitude 
$\langle\!\langle \{m\}|\Psi\rangle\!\rangle$ associated 
with the occupation numbers ${\{m\}\equiv [m_0,m_1,m_2,\ldots]}$ of the 
Landau orbitals as
\begin{equation}\label{eq:mps-amplitude}
\langle\!\langle \{m\}|\Psi\rangle\!\rangle
=\langle\mathrm{out}|
\cdots
\hat{C}^{m_2}
\hat{C}^{m_1}
\hat{C}^{m_0}
|\mathrm{in}\rangle,
\end{equation}
with the orbital $\hat{C}^m$ operators given by
\begin{equation}\label{eq:Bm-def}
\begin{aligned}
\hat{C}^0&=\hat{U}(\gamma),&
\hat{C}^1&=\hat{U}(\gamma)\,\hat{\mathcal{V}}_0.
\end{aligned}
\end{equation}
Upon choosing a basis for the CFT Hilbert space, the operator expression in 
Eq.~\eqref{eq:mps-amplitude} becomes a matrix product state.

\subsection{Cylinder evolution operator}\label{sec:cylinder-evolution}

We now go back to the issue of the uniform background charge.
We would like to treat it in the same way as the electron operator.
To this end, we first split the two-dimensional integral in 
$\mathcal{O}_\text{bc}$ [Eq.~\eqref{eq:bc}] into small patches, 
\begin{equation}
\!\!
:\!e^{-i\frac{\sqrt{\nu}}{2\pi}\int\!\mathrm{d}x\mathrm{d}y\,\hat{\phi}(x+iy)}\!:\,\,
\,\sim\prod_{x,y}
:\!e^{-i\frac{\sqrt{\nu}}{2\pi}\delta x\delta y\,\hat{\phi}(x+iy)}\!:,
\end{equation}
where the product over patches $(x,y)$ is time ordered, and $\nu$ is the 
filling fraction.
Evidently this operation introduces unwanted self-interactions between 
background charges at different locations. Fortunately, this only adds an 
overall constant factor that does not depend on the electron position.
Notice that each factor 
$:\!e^{-i\frac{\sqrt{\nu}}{2\pi}\delta x\delta y\,\hat{\phi}(x+iy)}\!:$
is now a primary field, to which Eq.~\eqref{eq:dilation} applies:
\begin{equation*}
:\!e^{-i\frac{\sqrt{\nu}}{2\pi}\delta x\delta y\,\hat{\phi}(x+iy)}\!:\,\,
=e^{x\gamma\hat{L}_0}
:\!e^{-i\frac{\sqrt{\nu}}{2\pi}\delta x\delta y\,\hat{\phi}(iy)}\!:
e^{-x\gamma\hat{L}_0}.
\end{equation*}
Thanks to the time ordering,
we can recombine the patches at the same $x$,
and up to an overall constant we have,
\begin{equation}
\!\!
\!\!
:\!e^{-i\frac{\sqrt{\nu}}{2\pi}\int\!\mathrm{d}x\mathrm{d}y\,\hat{\phi}(x+iy)}\!:\,\,
\sim\prod_{x}
e^{x\gamma\hat{L}_0}
e^{-i\frac{\sqrt{\nu}\delta x}{\gamma}\,\hat{\phi}_0}
e^{-x\gamma\hat{L}_0},
\end{equation}
where the product over time slices $x$ is still time ordered,
and the zero-mode $\hat{\phi}_0$ of the boson field 
[Eq.~\eqref{eq:boson-mode-expansion}] is picked up by
\begin{equation}
\hat{\phi}_0=\int_{-L_y/2}^{L_y/2}\frac{\mathrm{d}y}{L_y}\hat{\phi}(iy).
\end{equation}

Therefore, up to an inconsequential overall constant, the insertion of the 
uniform background charge operator amounts to injecting an exponentiated boson 
zero mode \emph{at each time slice}.
We can combine this with the time evolution, and redefine $\hat{U}(s)$ as 
the path-ordered exponential
\begin{equation}\label{eq:U-path}
\hat{U}(s)\equiv\mathcal{P}\exp\left[-\int_0^s
\mathrm{d}x\left(\gamma\hat{L}_0+i\frac{\sqrt{\nu}}{\gamma}\hat{\phi}_0\right)
\right],
\end{equation}
where $\hat{L}_0$ is the dilation operator [Eq.~\eqref{eq:L0-total}]
for the direct-product CFT.
This modification is enough to capture the effect of the uniform background 
charge operator.
Note that in Eq.~\eqref{eq:U-path}
the path dependence comes from the boson zero mode $\hat{\phi}_0$ 
and its canonical conjugate $\hat{a}_0$ hidden in $\hat{L}_0$ [Eq.~\eqref{eq:boson-L0}].
This allows us to simplify the path-ordered 
exponential,~\cite{Estienne13:MPSLong} yielding
\begin{multline}
\hat{U}(s)=\\
\exp\left(-i\frac{s\sqrt{\nu}}{\gamma}\hat{\phi}_0\right)
\exp\left[-s\gamma\hat{L}_0-\frac{s^2}{2}\left(
\sqrt{\nu}\hat{a}_0
+\frac{s\nu}{3\gamma}
\right)\!\right]\!,
\end{multline}
where the two exponentials do not commute.
This new expression for $\hat{U}(s)$ supersedes the original definition in 
Eq.~\eqref{eq:U-naive}, and it enters the second-quantized amplitude through 
the orbital operators $\hat{C}^0=\hat{U}(\gamma)$ and 
$\hat{C}^1=\hat{U}(\gamma)\hat{\mathcal{V}}_0$ [Eq.\eqref{eq:Bm-def}].

We emphasize that the above treatment of the cylinder evolution operator 
is not specific to correlators of the electron operator.
The resulting formula for $\hat{U}(s)$ applies generally to time-ordered 
correlators of any conformal primary fields in the presence of a background 
charge.
We will make use of this fact when we derive the MPS representation of the 
quasihole insertion in Section~\ref{sec:qh-abelian}.

Recall that $\hat{a}_0$ measures the U(1) charge in unit of 
$\sqrt{\nu}$ times the electron charge.
Using Eq.~\eqref{eq:boson-zero-mode}, we find that
\begin{equation}
[\hat{U}(s)]^{-1}\,\hat{a}_0\,\hat{U}(s)=\hat{a}_0
-\frac{s\sqrt{\nu}}{\gamma}.
\end{equation}
Letting $s$ be the orbital spacing $\gamma$, we see that the amount of 
background charge contributed by each Landau orbital is equal to $-\nu$ 
times the electron charge.
This indeed neutralizes the total electric charge at filling $\nu$.

\subsection{Matrix product factorization}\label{sec:matrix-product-factorization}

The second-quantized amplitude in Eq.~\eqref{eq:mps-amplitude}
can be readily converted into a matrix product state.
Between each pair of adjacent $\hat{C}^m$ operators, we can insert a unit 
resolution into a complete set of states over the conformal Hilbert space,
\begin{equation}\label{eq:CFT-unit-resolution}
\hat{\mathbb{I}}=\sum_\alpha |\alpha\rangle\langle\alpha|.
\end{equation}
For the free boson, the orthonormal basis states $|\alpha\rangle$ are 
simply normalized descendants under the U(1) current.
For the neutral CFT, the Virasoro descendants are in general not orthogonal, 
or even linearly independent. An orthonormal basis can be obtained through the
Gram-Schmidt process after eliminating the null modes.~\cite{Estienne13:MPS}

Due to the $e^{-\gamma^2\hat{L}_0}$ factor introduced by 
the cylinder evolution $\hat{U}(\gamma)$ in the $\hat{C}^m$ operators,
the CFT states with higher energy (as measured by $\gamma\hat{L}_0$) are 
exponentially suppressed at finite cylinder perimeter.
This allows us to truncate~\cite{Yurov90:TCSA}
the conformal Hilbert space by keeping only the 
lowest few levels of descendants. The resulting finite-dimensional vector 
space is the MPS auxiliary space, over which
the orbital $\hat{C}^m$ operators assume a matrix representation
\begin{equation}
[C^m]_{\alpha\beta}=\langle\alpha|\hat{C}^m|\beta\rangle.
\end{equation}
The calculation of these matrix elements is discussed in 
Appendix~\ref{sec:matrix-element}.
With the truncated representation of the $\hat{C}^m$ operators, the 
second-quantized amplitude in Eq.~\eqref{eq:mps-amplitude} becomes a product 
of matrices dotted into the boundary vectors, which can be evaluated 
numerically.~\cite{Estienne13:MPSLong}

\subsection{Abelian quasiholes}\label{sec:qh-abelian}

In the CFT formalism,~\cite{Moore91:MR}
similar to the electrons, a localized quasihole at 
$\eta\in\mathbb{C}$ is represented by a primary field insertion 
in the conformal correlator.
As a warm-up for the non-Abelian case, in the following we first discuss the 
Abelian quasihole, represented by
\begin{equation}\label{eq:Q-abelian}
\mathcal{Q}(\eta)=\,:\!e^{i\sqrt{\nu}\phi(\eta)}\!:.
\end{equation}
This operator couples only to the free boson, and it generates the familiar
quasihole ${\prod_i(z_i-\eta)}$ factor when evaluated in the plane (as opposed to the 
cylinder).
We now discuss how to add $\mathcal{Q}(\eta)$ to the MPS 
construction.~\cite{Zaletel12:MPS}

\begin{figure}[]
\centering
\includegraphics[]{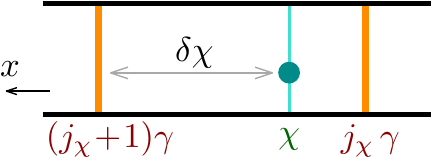}%
\caption{\label{fig:qh-insertion}
Insertion of a single quasihole at imaginary time $\chi$.
The solid vertical lines in orange mark the center positions of the two Landau 
orbitals sandwiching the quasihole.
}
\end{figure}

We consider a single Abelian quasihole at ${\eta=\chi+i\zeta}$.
In Eq.~\eqref{eq:operator-formalism}, we now have an extra 
$\hat{\mathcal{Q}}(\chi+i\zeta)$ operator inserted into the chain of electron 
operators, at the position determined by time ordering.
Similar to Eq.~\eqref{eq:dilation}, we can extract the $\chi$ dependence using 
the dilation operator,
\begin{equation}
\hat{\mathcal{Q}}(\chi+i\zeta)
=e^{\chi\gamma \hat{L}_0}\hat{\mathcal{Q}}(i\zeta)e^{-\chi\gamma\hat{L}_0}.
\end{equation}
Following the same steps from Sec.~\ref{sec:occupation-number}
to Sec.~\ref{sec:cylinder-evolution}, for each 
imaginary-time interval of size $s$ between adjacent primary field (electron 
or quasihole) insertions,
we can capture the dilation and the background charge using the
cylinder evolution operator $\hat{U}(s)$ [Eq.~\eqref{eq:U-path}].

As a result of the quasihole at $\chi+i\zeta$,
the cylinder evolution is now further punctured by $\hat{\mathcal{Q}}(i\zeta)$ 
at the time slice $\chi$.
To locate this time slice relative to the Landau orbitals,
recall that the orbital $j$ is centered at $x=\gamma j$ 
[Eq.~\eqref{eq:Landau-orbital}].
The $\hat{\mathcal{Q}}(i\zeta)$ operator is thus inserted between 
the orbitals $j_\chi$ and ${j_\chi+1}$ (Fig.~\ref{fig:qh-insertion}), with
\begin{equation}\label{eq:j-chi}
j_\chi\equiv\left\lfloor{\chi/\gamma}\right\rfloor.
\end{equation}
Here, the floor function $\left\lfloor{t}\right\rfloor\in\mathbb{Z}$
denotes the largest integer no greater than $t\in\mathbb{R}$.
The insertion of $\hat{\mathcal{Q}}(i\zeta)$
breaks the cylinder evolution $\hat{U}(\gamma)$ associated with 
the orbital $\hat{C}^m$ operator at $j_\chi$ into two parts,
\begin{equation}
\hat{U}(\gamma)\,\,\rightarrow\,\,
\hat{U}(\delta\chi)
\hat{\mathcal{Q}}(i\zeta)
\hat{U}(\gamma-\delta\chi),
\end{equation}
where $\delta\chi\in(0,\gamma]$ denotes the displacement of $\chi$ from the center of the 
orbital ${j_\chi+1}$ (Fig.~\ref{fig:qh-insertion}),
\begin{equation}\label{eq:delta-chi}
\delta\chi\equiv (j_\chi+1)\gamma-\chi.
\end{equation}
Note that the cylinder evolution operator $\hat{U}(s)$ does not commute with the 
$\hat{\mathcal{Q}}(i\zeta)$ insertion, and from its definition in term of the 
path-ordered exponential in Eq.~\eqref{eq:U-path}, we have
\begin{equation}
\hat{U}(\gamma-\delta\chi)
=[\hat{U}(\delta\chi)]^{-1}\,\,
\hat{U}(\gamma).
\end{equation}
Then, without modifying the $\hat{C}^m$ operators, the quasihole 
$\hat{\mathcal{Q}}(\chi+i\zeta)$ can be represented in 
Eq.~\eqref{eq:mps-amplitude} by the insertion of
\begin{equation}\label{eq:qh-without-sign}
\hat{U}(\delta\chi)\,
\hat{\mathcal{Q}}(i\zeta)\,
[\hat{U}(\delta\chi)]^{-1}
\end{equation}
between the $\hat{C}^m$
operators for orbitals $j_\chi$ and $j_\chi+1$.

There is one extra complication that we have glossed over.
When we expand the CFT-derived wave function
\begin{equation}
\Big\langle
\mathcal{V}(z_1)\cdots \mathcal{V}(z_n)\,\mathcal{Q}(\chi+i\zeta)\,\,
\mathcal{O}_\text{bc}
\Big\rangle
\end{equation}
into Slater-determinant basis 
states in Eq.~\eqref{eq:correlator-psi-j}, we place the electron contours 
at the center of each occupied orbital.
To bring the field insertions into time ordering, the electrons on orbitals 
with center position $x<\chi$ have to be moved across the pinned quasihole 
field at time $\chi$.
For the Abelian quasihole as in Eq.~\eqref{eq:Q-abelian}, each of these 
commutations incurs a minus sign,
\begin{equation}\label{eq:anticommute-abelian}
\langle
\cdots \mathcal{V}(z)\mathcal{Q}(\chi+i\zeta) \cdots
\rangle
= -\,
\langle
\cdots \mathcal{Q}(\chi+i\zeta)\mathcal{V}(z) \cdots
\rangle.
\end{equation}
The above anti-commutativity simply reflects the fact that in the planar wave 
function an Abelian quasihole is represented by the \emph{odd}-power factor 
${\prod_i (w_i-\eta)}$.
Formally, one could also derive this minus sign from the operator product 
expansion.

We need to collect these minus signs together and attach them to the quasihole 
insertion.
To this end, for each Slater-determinant basis state we have to count the 
number of occupied orbitals with center position $x<\chi$.
This number can be extracted from the conserved U(1) charge at the time slice 
$\chi$ of the quasihole insertion.
[Note that for each orbital the cylinder evolution $\hat{U}(\gamma)$ contained 
in the $\hat{C}^m$ operator correctly accounts for the associated background 
charge.]
Specifically, the number of occupied orbitals with $x<\chi$ is given by
\begin{equation}
\sqrt{\nu}\hat{a}_0+(j_\chi+1)\nu
\end{equation}
inserted between the $\hat{C}^m$ operators for orbitals $j_\chi$ and 
$j_{\chi}+1$.
Here, the zero-mode $\hat{a}_0$ measures the U(1) charge in unit of 
$\sqrt{\nu}$ times the electron charge $e$, and the second term 
cancels the background charge $-\nu e$ carried by each orbital
$j\in\{0,1,\ldots,j_\chi\}$.
Finally, we can write down the full expression for the quasihole operator in 
the MPS
\begin{equation}\label{eq:B-Abelian-qh}
\hat{U}(\delta\chi)\,
\hat{\mathcal{Q}}(i\zeta)\,
[\hat{U}(\delta\chi)]^{-1}\,
(-1)^{\sqrt{\nu}\hat{a}_0+(j_\chi+1)\nu}.
\end{equation}
Note that $\hat{a}_0$ does not commute with $[\hat{U}(\delta\chi)]^{-1}$, 
as the latter contains $\hat{\phi}_0$.

The above construction can be easily generalized to the case of multiple 
Abelian quasiholes, with
\begin{equation}\label{eq:multiple-qh}
\Big\langle
\mathcal{V}(z_1)\cdots \mathcal{V}(z_n)\,
\mathcal{Q}(\chi_1+i\zeta_1)\cdots\mathcal{Q}(\chi_m+i\zeta_m)\,
\mathcal{O}_\text{bc}
\Big\rangle.
\end{equation}
We put the $m$ quasiholes in time ordering,
\begin{equation}\label{eq:qh-time-ordering}
\chi_1>\chi_2>\cdots>\chi_m.
\end{equation}
Then, each quasihole has the MPS representation given by 
Eq.~\eqref{eq:B-Abelian-qh}, except for one minor modification.
For the ${l\text{-th}}$ quasihole, to extract the number of occupied orbitals with 
$x<\chi_l$, we need to subtract not only the background charge, but also the 
U(1) charge introduced by the other quasiholes inserted on its right.
This modifies the commutation sign in Eq.~\eqref{eq:B-Abelian-qh} to
\begin{equation}\label{eq:abelian-qh-sign}
(-1)^{\sqrt{\nu}\hat{a}_0+(j_\chi+1)\nu-(m-l)\nu}.
\end{equation}
The last two terms in the exponent [$(j_\chi+1)\nu-(m-l)\nu$]
only introduce an overall constant phase factor (independent from the electron
occupations), but they are necessary to eliminate the branch-cut ambiguity in 
the fractional power $(-1)^{\sqrt{\nu}\hat{a}_0}$.
This branch cut ambiguity comes from the fact that the operator 
$\sqrt{\nu} a_0$ now counts fractional quasihole charges.

This finishes our review of the MPS construction in the absence of non-Abelian 
quasiholes. We conclude this section with a few remarks.
First, the quasihole MPS has a rather subtle parametric dependence on the 
quasihole position $\chi+i\zeta$.
This dependence is not holomorphic due to the presence of the nonchiral 
background charge operator $\mathcal{O}_\text{bc}$, and the $\chi$-dependence
differs considerably from $\zeta$.
The imaginary part $\zeta$ enters the MPS directly (and solely) through the 
operator insertion $\hat{\mathcal{Q}}(i\zeta)$.
The real part $\chi$ controls the location $j_\chi$ of this insertion, and 
thereby affects both the cylinder evolution and the electron-quasihole 
commutation sign.
Second, we note that the amount of U(1) charge carried by an Abelian quasihole 
operator $\hat{\mathcal{Q}}(i\zeta)$ is exactly opposite to the charge carried 
by an empty orbital $\hat{C}^0=\hat{U}(\gamma)$.
This can be seen by comparing the exponents in Eqs.~\eqref{eq:Q-abelian} 
and~\eqref{eq:U-path}.
Therefore, to keep the CFT boundary states fixed, upon inserting a quasihole
we need to increase the total number of Landau orbitals by one.
Third, in the presence of multiple quasiholes, the conformal correlator 
exhibits a nontrivial monodromy as a function of quasihole positions.
In the MPS, this monodromy manifests itself as branch cuts originating from 
the center of each quasihole. We will discuss this in details in 
Sec.~\ref{sec:branch-cut}.
Finally, we emphasize that our prescription for the Abelian quasihole matrix 
differs from Ref.~\onlinecite{Zaletel12:MPS} in the more careful handling of 
the background 
charge. The new formula here exactly preserves the quasihole-dependent 
normalization of the conformal correlator.
This feature is desirable as it enables us to leverage the plasma 
analogy when checking the braiding statistics.~\cite{Wu14:Quasihole}

\section{Non-Abelian Quasiholes}
\label{sec:non-Abelian}

We now proceed to the construction of MPS for the non-Abelian quasiholes.
We consider the so-called $(k,r)$-clustered states~\cite{Bernevig08:Jack}
at filling fraction $\nu$ of the lowest Landau level, with $k,r$ being 
integers and
\begin{equation}
\nu=\frac{k}{k+r}.
\end{equation}
The electronic correlations in such states are characterized by the presence 
of $k$-particle clusters.
The fermionic wave function is given by the product of a Jastrow factor and a 
bosonic part.
The latter bosonic wave function vanishes to order $r$
when $(k+1)$ particles come together.
The $r=2$ case corresponds to the $\mathbb{Z}_k$ Read-Rezayi 
series,~\cite{Read99:RR} with $k=2$ being the Moore-Read 
state,~\cite{Moore91:MR} while the $(k,r)=(2,3)$ case is the so-called 
Gaffnian wave function.~\cite{Simon07:Gaffnian}
These wave functions can be constructed from chiral conformal correlators of 
primary fields representing electrons and quasiholes.~\cite{Read99:RR}
The electron operator takes the tensor product form
\begin{equation}\label{eq:Ve}
\mathcal{V}(z)=\psi(z)\,\otimes :\!e^{i\frac{1}{\sqrt{\nu}}\phi(z)}\!:,
\end{equation}
where in addition to the free-boson vertex operator,
we also have a primary field $\psi$ in the so-called neutral CFT.
The fundamental quasihole is represented by
\begin{equation}\label{eq:qh-fundamental}
\mathcal{Q}(\eta)=\sigma(\eta)\,\otimes
:\!e^{i\frac{\sqrt{\nu}}{k}\phi(\eta)}\!:,
\end{equation}
with $\sigma$ being another primary field in the neutral CFT.
The reduced exponent in the boson vertex operator reflects the fact that the 
fundamental quasihole is a further $k$-fold fractionalization of the 
Abelian quasihole in Eq.~\eqref{eq:Q-abelian}.

The MPS auxiliary space is now given by the direct product
of the truncated Hilbert spaces of the neutral CFT and the free boson.
As noted in Sec.~\ref{sec:review}, the free boson Hilbert space can be 
naturally broken into sectors labeled by the conserved U(1) charge $\hat{a}_0$.
The neutral CFT Hilbert space can be similarly split into different 
representations of the neutral Virasoro algebra.~\cite{DiFrancesco99:Yellow}
Each representation, called a Verma module, is spanned by the 
conformal family of Virasoro descendants generated from a single primary state. 
Therefore, each Verma module of the neutral CFT is labeled by 
a primary field, which we refer to as the ``topological charge'' and 
denote by Latin indices $a,b,c,\ldots$.
Taken together, the MPS auxiliary space can be split into different sectors 
labeled by the topological and the U(1) charges.~\cite{Estienne13:MPSLong}
For later convenience, we define the projector into a single Verma module $c$ as
\begin{equation}\label{eq:Verma-projector}
\hat{\mathcal{P}}(c)=\sum_{\alpha\in c}
|\alpha\rangle\langle\alpha|.
\end{equation}

As shown in Sec.~\ref{sec:qh-abelian},
a quasihole at $\chi+i\zeta$ is represented by the insertion of 
$\hat{\mathcal{Q}}(i\zeta)$ into the cylinder evolution at time slice $\chi$,
as in Eq.~\eqref{eq:qh-without-sign}.
There are two extra complications for the non-Abelian case in 
Eq.~\eqref{eq:qh-fundamental}.
First, we need to resolve the topologically degenerate states associated with 
multiple pinned non-Abelian quasiholes.
Second, we need to generalize the anti-commutativity between electrons and 
Abelian quasiholes in Eq.~\eqref{eq:anticommute-abelian}.
As noted earlier, a non-Abelian quasihole can be seen as a further $k$-fold 
fractionalization of an Abelian quasihole.
This hints at a $k$-way split of the anti-commutation minus sign. 
However, to have a single-valued electron wave function, the commutation 
phase between an electron and a quasihole must square to unity.
As we demonstrate below, the solution to this conundrum turns out to be 
letting each of the $k$ parts of a quasihole anticommute
with \emph{only one out of every $k$ electrons}.

\subsection{Neutral CFT examples}

Before diving into the details of the quasihole operator,
we first go through the field content and the fusion rules of the 
neutral CFT for a few representative theories.~\cite{Estienne13:MPS,Estienne13:MPSLong}

\emph{Moore-Read.}
The neutral CFT for the Moore-Read Pfaffian state~\cite{Moore91:MR}
is the minimal model $\mathcal{M}(3,4)$ with central charge $c=\frac{1}{2}$.
The primary fields $(\mathbbm{1},\psi,\sigma)$ 
have scaling dimensions $(0,\frac{1}{2},\frac{1}{16})$.
The fusion rules are given by
\begin{equation}
\begin{aligned}
\psi\times\psi&=\mathbbm{1},&
\sigma\times\psi&=\sigma,&
\sigma\times\sigma&=\mathbbm{1}+\psi.
\end{aligned}
\end{equation}

\emph{Gaffnian.}
The neutral CFT for the Gaffnian wave function~\cite{Simon07:Gaffnian}
is the nonunitary minimal 
model $\mathcal{M}(3,5)$, with a negative central charge $c=-\frac{3}{5}$.
The primary fields $(\mathbbm{1},\psi,\sigma,\varphi)$ have scaling dimensions
$(0,\frac{3}{4},-\frac{1}{20},\frac{1}{5})$.
The fusion rules involving $\psi$ or $\sigma$ are given by
\begin{equation}
\begin{aligned}
\psi\times\psi&=\mathbbm{1},&
\psi\times\sigma&=\varphi,&
\psi\times\varphi&=\sigma,\\
\sigma\times\sigma&=\mathbbm{1}+\varphi,&
\sigma\times\varphi&=\psi+\sigma.
\end{aligned}
\end{equation}

\emph{$\mathbb{Z}_3$ Read-Rezayi.}
The neutral CFT for the $\mathbb{Z}_3$ Read-Rezayi state is the $\mathbb{Z}_3$ 
parafermionic variant~\cite{Read99:RR}
of the ${c=\frac{4}{5}}$ minimal model $\mathcal{M}(5,6)$.
The Virasoro primary fields
$(\mathbbm{1},\psi_1,\psi_2,W,\varepsilon,\sigma_1,\sigma_2,\varphi)$
have scaling dimensions 
$(0,\frac{2}{3},\frac{2}{3},3,\frac{2}{5},\frac{1}{15},\frac{1}{15},\frac{7}{5})$
and $\mathbb{Z}_3$ charges ${(0,2,1,0,0,1,2,0)}$.
This theory actually enjoys an extended $\mathcal{W}_3$ algebra~\cite{Estienne09:W}
beyond the Virasoro.
In this language, the $W$ field is not a primary field, but rather a 
descendant of the identity under the larger $\mathcal{W}_3$ algebra.
Similarly, $\varphi$ is actually a descendant of $\varepsilon$.
However, as noted in Ref.~\onlinecite{Estienne13:MPSLong}, the $\mathcal{W}_3$ 
approach is numerically inefficient due to the complexity of the extended 
algebra and the proliferation of null modes.
Hence, in the following we stick to the Virasoro description,
but for succinctness we keep $W$ and $\varphi$ implicit whenever possible.
With this caveat in mind, the fusion rules involving the electron 
$\psi\equiv\psi_1$ or the quasihole $\sigma\equiv\sigma_1$ are 
given by
\begin{equation*}
\begin{minipage}[h]{0.95\linewidth}
\begin{ruledtabular}
\begin{tabular}{c|cccccc}
& $\psi_1$ & $\psi_2$ & $\varepsilon$ & $\sigma_1$ & $\sigma_2$ \\\hline\hline
$\psi_1$ & $\psi_2$ & $\mathbbm{1}$ & $\sigma_2$ & $\varepsilon$ & $\sigma_1$ \\\hline
$\sigma_1$ & $\varepsilon$ & $\sigma_2$ & $\psi_2+\sigma_1$ & $\psi_1+\sigma_2$ & $\mathbbm{1}+\varepsilon$
\end{tabular}
\end{ruledtabular}
\end{minipage}
\end{equation*}

Note that the $\sigma\times\sigma$ fusion has multiple outcomes in all of the 
examples above.
Such fusion ambiguity is characteristic of the CFT representation of the 
non-Abelian quasiholes.
For completeness, we list the structure constants for the so-called operator product expansion needed to construct the MPS matrix elements in 
Appendix~\ref{sec:ope}.

\subsection{Conformal blocks and fusion trees}\label{sec:fusion-trees}

We consider conformal correlators with multiple non-Abelian quasihole insertions
\begin{equation}\label{eq:multiple-qh-nonAbelian}
\Big\langle
\mathcal{V}(z_1)\cdots \mathcal{V}(z_n)\,
\mathcal{Q}(\eta_1)\cdots\mathcal{Q}(\eta_m)\,\,
\mathcal{O}_\text{bc}
\Big\rangle.
\end{equation}
Due to the nontrivial fusion rule of the $\sigma$ field in $\mathcal{Q}(\eta)$ 
[Eq.~\eqref{eq:qh-fundamental}], for each set of quasihole coordinates, the 
above expression does not produce a single wave function.
Instead, it defines a vector space of degenerate wave functions.~\cite{Moore91:MR}
A set of basis states in this space, called conformal blocks, can be 
obtained by specifying how the fields fuse together in terms of a fusion tree 
diagram.~\cite{Moore89:CFT,DiFrancesco99:Yellow}
We only need to consider the neutral CFT, since the free boson has 
trivial fusion rules dictated by the U(1) charge conservation.
The structure of the fusion tree reflects the ordering of the successive 
fusions, and different structures correspond to different basis choices for
the same vector space.
For a given fusion tree structure, the corresponding conformal-block basis states
can be enumerated by finding all the topological charge labelings compatible with 
the fusion rules.

We now consider the MPS representation of the conformal blocks associated with 
Eq.~\eqref{eq:multiple-qh-nonAbelian}.
This mandates the fields to be fused \emph{sequentially} in time ordering into 
the $|\text{in}\rangle$ state.
As explained in Sec.~\ref{sec:qh-abelian}, each quasihole is pinned at a 
particular time slice while the electrons need to be placed at the center of 
each occupied orbital. This requires us to blend the quasihole operators into
the chain of electron operators.
Specifically, we have a $\hat{\mathcal{V}}_0$ operator placed at the center of 
each occupied Landau orbital and a $\hat{\mathcal{Q}}(i\zeta)$ operator 
inserted at each quasihole position $\chi$,
and the operators are lined up in time ordering.
This leads to a fusion tree with a \emph{linear} structure.
Consider for example the following amplitude for the Moore-Read state
\begin{equation}\label{eq:amplitude-example}
\langle
\cdots
\hat{\mathcal{V}}_0\,
\hat{\mathcal{Q}}(i\zeta_1)\,
\hat{\mathcal{V}}_0\,
\hat{\mathcal{Q}}(i\zeta_2)\,
|\mathbbm{1}\rangle.
\end{equation}
To reduce clutter, here we have omitted the interleaving cylinder evolution 
operators $\hat{U}(s)$.
The fusion tree takes the form
\begin{equation}\label{eq:primitive-tree}
\begin{minipage}[c]{2.5cm}
\includegraphics[width=\linewidth]{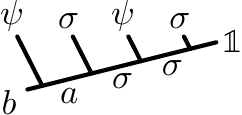}
\end{minipage}
\end{equation}
where the undecided topological charges $(a,b)$ could be either 
$(\mathbbm{1},\psi)$ or $(\psi,\mathbbm{1})$ according to the fusion rules.
It should be noted that the imaginary time $x$ points in the \emph{left} 
direction in the above diagram, in accordance with the operator time ordering.
To construct the conformal block for either choice of the topological charges 
$(a,b)$, we need to materialize the 
fusion channel choice in Eq.~\eqref{eq:amplitude-example}.
This amounts to inserting a Verma module projector to the left of each field 
insertion,
\begin{equation}\label{eq:amplitude-example-resolved}
\langle
\cdots
\hat{\mathcal{P}}(b)\,\hat{\mathcal{V}}_0\,
\hat{\mathcal{P}}(a)\,\hat{\mathcal{Q}}(i\zeta_1)\,
\hat{\mathcal{P}}(\sigma)\,\hat{\mathcal{V}}_0\,
\hat{\mathcal{P}}(\sigma)\,\hat{\mathcal{Q}}(i\zeta_2)\,
|\mathbbm{1}\rangle.
\end{equation}

It should be noted that the placement of the electron $\hat{\mathcal{V}}_0$ operators
depends on how and where the Landau orbitals are occupied.
As a result, with quasihole insertions in the bulk, the fusion trees for 
different Slater-determinant amplitudes naturally have \emph{different} 
orderings of their quasihole and electron ``branches''.
For example, with a non-Abelian quasihole pinned at $\eta=\frac{3}{2}\gamma$,
the following two Slater-determinant amplitudes have different operator 
ordering and thereby different fusion tree structure,
\begin{equation}
\begin{aligned}
\cdots \mathtt{101010}_{\!{}_{\star\!}\!}\mathtt{11}&
& \Rightarrow\quad \langle\cdots
&\hat{\mathcal{V}}_0\hat{\mathcal{V}}_0\hat{\mathcal{V}}_0
\hat{\mathcal{Q}}(0)
\hat{\mathcal{V}}_0
\hat{\mathcal{V}}_0|\mathbbm{1}\rangle, \\
\cdots \mathtt{011011}_{\!{}_{\star\!}\!}\mathtt{01}&
& \Rightarrow\quad \langle\cdots
&\hat{\mathcal{V}}_0\hat{\mathcal{V}}_0\hat{\mathcal{V}}_0
\hat{\mathcal{V}}_0
\hat{\mathcal{Q}}(0)
\hat{\mathcal{V}}_0|\mathbbm{1}\rangle.
\end{aligned}
\end{equation}
Here the occupation numbers are listed in reverse to be consistent with the 
time ordering, the star marks the position of the quasihole relative to the 
Landau orbitals, and again we have omitted the interleaving cylinder 
evolutions in order to highlight the operator ordering.
Now, to obtain the correct wave function, we must resolve \emph{every}
Slater-determinant amplitude in the \emph{same} fusion tree basis.
A natural choice is to \emph{have all the quasihole fields placed at the beginning 
(rightmost end) of the linear structure}. For the example given above, we want
\begin{equation}\label{eq:reordered-tree}
\begin{minipage}[c]{2.5cm}
\includegraphics[width=\linewidth]{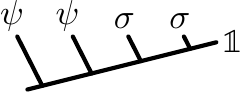}
\end{minipage}.
\end{equation}
For each Slater-determinant amplitude, we need to reorganize the time-ordered 
fusion tree in Eq.~\eqref{eq:primitive-tree}, by bringing all the quasihole 
branches across the electrons to the rightmost, while keeping their relative 
ordering.
As we show next, this reshuffling transformation amounts to adding a 
particular minus sign to each quasihole operator, 
generalizing the prescription for 
the Abelian quasihole in Eq.~\eqref{eq:abelian-qh-sign}.

\subsection{Exchanging two branches}

To connect the fusion tree with the desired structure in 
Eq.~\eqref{eq:reordered-tree} to the primitive time-ordered tree in 
Eq.~\eqref{eq:primitive-tree}, we need to move the 
quasihole insertions across electron operators into the bulk.
The elementary move is
\begin{equation}
\begin{minipage}[c]{1.5cm}
\includegraphics[width=\linewidth]{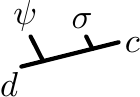}
\end{minipage}                                              
\quad\Rightarrow\quad                                       
\begin{minipage}[c]{1.5cm}                                  
\includegraphics[width=\linewidth]{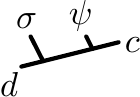}
\end{minipage}.
\end{equation}
These two fusion trees span the same vector space, and they can be related 
by a linear transform
\begin{equation}\label{eq:half-braid-swap}
\begin{minipage}[c]{1.5cm}
\includegraphics[width=\linewidth]{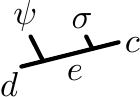}
\end{minipage}
=\sum_f \big[B^{\sigma\psi c}_d\big]_{ef}
\begin{minipage}[c]{1.5cm}
\includegraphics[width=\linewidth]{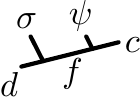}
\end{minipage},
\end{equation}
which is nothing but the CFT half-braid matrix between the electron and the 
quasihole fields.
It should be noted that the U(1) part of the CFT also contributes a phase 
factor, even though it has trivial fusions and has been omitted from the 
above diagrams.
Also, for the CFT correlator to represent a physical wave function, the 
electron operator must be local with respect to the quasiholes.
As a result, it does not matter whether the electron-quasihole half braid is 
done clockwise or counterclockwise.
This brings Eq.~\eqref{eq:half-braid-swap} to a more familiar form,
\begin{equation}\label{eq:half-braid-familiar}
\begin{minipage}[c]{1.5cm}
\includegraphics[width=\linewidth]{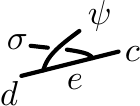}
\end{minipage}
=\sum_f \big[B^{\sigma\psi c}_d\big]_{ef}
\begin{minipage}[c]{1.5cm}
\includegraphics[width=\linewidth]{B_move_right.pdf}
\end{minipage}.
\end{equation}

Formally, the half-braid matrix $B^{abc}_d$
can be decomposed into the so-called $F$ and 
$R$ moves of the direct-product CFT.~\cite{Moore89:CFT}
Again, the contribution from the U(1) part must be included despite its 
omission from the diagrams.
The fusion $F$ matrix is a generalization of the Wigner $6j$-symbol,
\begin{equation}
\begin{minipage}[c]{1.5cm}
\includegraphics[width=\linewidth]{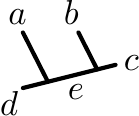}
\end{minipage}
=\sum_g \big[F^{abc}_d\big]_{eg}
\begin{minipage}[c]{1.5cm}
\includegraphics[width=\linewidth]{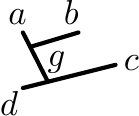}
\end{minipage},
\end{equation}
where $g$ is summed over all topological charges compatible with the fusion 
rules.
And the $R$ matrix gives the exchange phase in a definite fusion channel,
\begin{equation}
\begin{minipage}[c]{0.75cm}
\includegraphics[width=\linewidth]{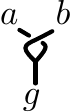}
\end{minipage}
=R^{ba}_g
\begin{minipage}[c]{0.75cm}
\includegraphics[width=\linewidth]{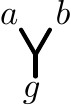}
\end{minipage}.
\end{equation}
Note that the fusion tree layout in the above definitions is slightly 
different from the standard convention in the literature.~\cite{Kitaev06:Anyon}
This is due to our choice of pointing the imaginary time $x$ in the \emph{left} 
direction in accordance with the operator time ordering.
Composing the $F$ and $R$ moves, we find
\begin{equation}
\begin{minipage}[c]{1.5cm}
\includegraphics[width=\linewidth]{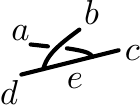}
\end{minipage}
=\sum_{f,g}
\big[F^{bac}_d\big]_{eg}\,
R^{ba}_g
\Big[\big(F^{abc}_d\big)^{-1}\Big]_{gf}
\begin{minipage}[c]{1.5cm}
\includegraphics[width=\linewidth]{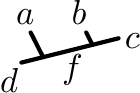}
\end{minipage}.
\end{equation}
The transformation coefficient in the above equation is nothing but 
$\big[B^{abc}_d\big]_{ef}$.
In principle, we can solve the pentagon and the hexagon equations~\cite{Moore89:CFT}
for the $F$ and the $R$ matrices, and fix the gauge according to the structure 
constants in the operator product expansion of the chiral fields (see 
Appendix~\ref{sec:ope}).
This would be a time-consuming task.
In practice, however, we can more easily determine the $B^{\sigma\psi c}_d$ 
matrix from a simple numerical calculation.
To this end, we consider the conformal-block version of Eq.~\eqref{eq:half-braid-swap},
\begin{multline}\label{eq:B-conformal-block}
\langle d|\,
\hat{\mathcal{V}}(iy)\,
\hat{\mathcal{P}}(e)\,
\hat{\mathcal{Q}}(i\zeta)\,
|c\rangle=\\
\sum_f \big[B^{\sigma\psi c}_d\big]_{ef}
\langle d|\,
\hat{\mathcal{Q}}(i\zeta)\,
\hat{\mathcal{P}}(f)\,
\hat{\mathcal{V}}(iy)\,
|c\rangle.
\end{multline}
Here, we place the operators at adjacent time slices to avoid complications 
from the cylinder evolution, and for simplicity we choose the end states 
$\langle d|$ and $|c\rangle$ to be the primary states in the 
corresponding Verma modules.
Recall that the core part of the MPS implementation is a truncated CFT calculator.
By numerically evaluating the two (truncated) correlators in the above 
equation as functions of $(y,\zeta)$, we can extract the $B^{\sigma\psi c}_d$ 
matrix in the gauge determined by the chiral structure constants.

We first consider the Moore-Read and the Gaffnian states.
They are clustered wave functions with ${k=2}$.
For these theories, the fusion $\psi\times a$ always yields a unique result 
for any topological charge $a$.
Therefore, the topological charges $d$ and $f$ in 
Eq.~\eqref{eq:half-braid-swap} are completely fixed by the choice of $c$ and 
$e$, namely
\begin{equation}
\begin{aligned}
d&=\psi\times e,&
f&=\psi\times c,
\end{aligned}
\end{equation}
with $e$ among the (possibly) multiple fusion outcomes of $\sigma\times c$.
This allows us to adopt the shorthand notation
\begin{equation}\label{eq:B-shorthand}
\big[B^{\sigma\psi c}_d\big]_{ef}
\,\,\rightarrow\,\,\,
B^c_e,
\end{equation}
and it reduces Eq.~\eqref{eq:B-conformal-block} to
\begin{multline}\label{eq:B-conformal-block-shorthand}
\langle \psi\times e|\,
\hat{\mathcal{V}}(iy)\,
\hat{\mathcal{P}}(e)\,
\hat{\mathcal{Q}}(i\zeta)\,
|c\rangle=\\
B^c_e\,\cdot\,
\langle \psi\times e|\,
\hat{\mathcal{Q}}(i\zeta)\,
\hat{\mathcal{P}}(\psi\times c)\,
\hat{\mathcal{V}}(iy)\,
|c\rangle.
\end{multline}
The single-valuedness of the electron wave function dictates that $B^c_e$ has 
to be $\pm 1$.
For the Moore-Read state, from the numerics we find
\begin{equation}\label{eq:B-move-MR}
\begin{aligned}
B^\mathbbm{1}_\sigma&=B^\sigma_\psi=1,&
B^\psi_\sigma&=B^\sigma_\mathbbm{1}=-1.
\end{aligned}
\end{equation}
And for the Gaffnian state, we have
\begin{equation}\label{eq:B-move-Gf}
B^\mathbbm{1}_\sigma=B^\varphi_\psi=B^\sigma_\varphi=B^\varphi_\sigma=1,\quad
B^\sigma_\mathbbm{1}=B^\psi_\varphi=-1.
\end{equation}
Depending on the fusion channel context, the electron and the quasihole fields 
either commute or anticommute.

For the $\mathbb{Z}_{k=3}$ Read-Rezayi state, due to our choice of keeping the 
$\mathcal{W}_3$ algebra implicit and treating $W$ and $\varphi$ as primary 
fields, even the electron field $\psi\equiv\psi_1$ has nontrivial fusion rules,
\begin{equation}
\begin{aligned}
\psi_1\times\psi_2&=\mathbbm{1}+W,&
\psi_1\times\sigma_1&=\varepsilon+\varphi,
\end{aligned}
\end{equation}
which would not bode well for our proposed method.
However, the implicit $\mathcal{W}_3$ symmetry forbids us to split the 
channels on the right-hand side into separate conformal blocks.
To take advantage of this, we bind the Verma module $W$ to $\mathbbm{1}$ and 
bind $\varphi$ to $\varepsilon$ by redefining the projectors
[Eq.~\eqref{eq:Verma-projector}],
\begin{equation}
\begin{aligned}
\hat{\mathcal{P}}(\mathbbm{1})&\equiv
\sum_{\alpha\in \mathbbm{1},W}
|\alpha\rangle\langle\alpha|,&
\hat{\mathcal{P}}(\varepsilon)&\equiv
\sum_{\alpha\in \varepsilon,\varphi}
|\alpha\rangle\langle\alpha|,
\end{aligned}
\end{equation}
and then we simply forget about $W$ and $\varphi$ when labeling the fusion 
trees.
Now that $W$ and $\varphi$ are formally gone, we are allowed to use the 
shorthand notation in Eq.~\eqref{eq:B-shorthand}.
Crucially, we verify from the numerics that after such patching (but not before), 
the reduced linear transform in Eq.~\eqref{eq:B-conformal-block-shorthand} 
still holds.
This compatibility can be attributed to the underlying extended $\mathcal{W}_3$ 
algebra.
The $B^c_e$ coefficients are found to be
\begin{equation}\label{eq:B-move-RR}
\begin{aligned}
B^\mathbbm{1}_{\sigma_1}&=B^{\psi_1}_\varepsilon
=B^\varepsilon_{\psi_2}
=B^{\sigma_1}_{\psi_1}=B^{\sigma_1}_{\sigma_2}
=B^{\sigma_2}_\varepsilon=1,\\
B^{\psi_2}_{\sigma_2}&=B^\varepsilon_{\sigma_1}=B^{\sigma_2}_\mathbbm{1}=-1.
\end{aligned}
\end{equation}

\subsection{Reshuffling quasiholes into the bulk}

Through successive applications of the elementary move in 
Eq.~\eqref{eq:half-braid-swap}, we can achieve the reshuffling
\begin{equation}\label{eq:successive-shuffling}
\begin{minipage}[c]{2.5cm}
\includegraphics[width=\linewidth]{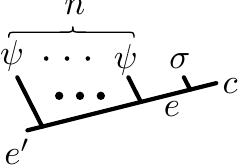}
\end{minipage}
\quad\Rightarrow\quad
\begin{minipage}[c]{2.5cm}
\includegraphics[width=\linewidth]{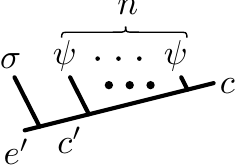}
\end{minipage}
\end{equation}
across an arbitrary number $n$ of electrons.
Note that the topological charges at the top and the bottom
must agree between the two trees;
otherwise the conformal blocks belong to distinct linear spaces.
Also, due to the trivial fusion rules of the $\psi$ field in the Moore-Read,
the Gaffnian, and the $\mathbb{Z}_3$ Read-Rezayi states,
all the topological charges are completely fixed by $c$ and $e$.
In particular, we deduce that the \emph{in-situ} fusion channels for the 
$\sigma$ field after the reshuffling are given by the successive fusion with 
$n$ $\psi$ fields,
\begin{equation}
\begin{aligned}
e'&=\psi^n\times e,&
c'&=\psi^n\times c,
\end{aligned}
\end{equation}
and the corresponding projected quasihole operator is
\begin{equation}\label{eq:qh-after-reshuffling}
\hat{\mathcal{Q}}(e,c,n,i\zeta)\equiv
\hat{\mathcal{P}}\big(\psi^n\times e\big)\,
\hat{\mathcal{Q}}(i\zeta)\,\,
\hat{\mathcal{P}}\big(\psi^n\times c\big).
\end{equation}

The linear transform between the two trees in 
Eq.~\eqref{eq:successive-shuffling} consists of a simple sign factor,
\begin{equation}\label{eq:successive-shuffling-sign}
\begin{minipage}[c]{2.5cm}
\includegraphics[width=\linewidth]{shuffle_n_left.pdf}
\end{minipage}
=(-1)^{N(e,c,n)}\,\cdot\,\,
\begin{minipage}[c]{2.5cm}
\includegraphics[width=\linewidth]{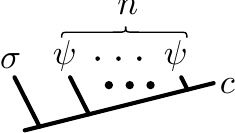}
\end{minipage}.
\end{equation}
Here, only a subset of the $n$ $\psi$ fields anticommute with $\sigma$, and
\begin{equation}
N(e,c,n)\in\mathbb{Z}
\end{equation}
counts the size of this anticommuting subset.
Consider $(e,c,n)=(\sigma,\mathbbm{1},4)$ for the Moore-Read state as an 
example. Recall from Eq.~\eqref{eq:B-move-MR} that $B^\mathbbm{1}_\sigma=1$ and 
$B^\psi_\sigma=-1$.
Using Eq.~\eqref{eq:half-braid-swap}, we find
\begin{equation}
\begin{aligned}
\begin{minipage}[c]{2.9cm}
\includegraphics[width=\linewidth]{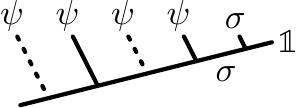}
\end{minipage}
&=B^\mathbbm{1}_\sigma
\begin{minipage}[c]{2.9cm}
\includegraphics[width=\linewidth]{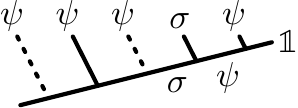}
\end{minipage}\\
&=B^\mathbbm{1}_\sigma B^\psi_\sigma
\begin{minipage}[c]{2.9cm}
\includegraphics[width=\linewidth]{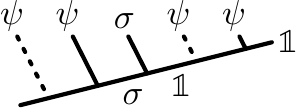}
\end{minipage}\\
&=B^\mathbbm{1}_\sigma B^\psi_\sigma B^\mathbbm{1}_\sigma B^\psi_\sigma
\begin{minipage}[c]{2.9cm}
\includegraphics[width=\linewidth]{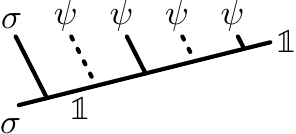}
\end{minipage}.
\end{aligned}
\end{equation}
In the above commuting process, the $\sigma$ field picks up a minus sign
$B^\psi_\sigma=-1$ from only \emph{every other} $\psi$ field (dashed line).
This is a direct consequence of Eq.~\eqref{eq:B-move-MR}.
The alternating pattern in the chain of $\psi$ fields has periodicity two, 
as expected from the fusion rule $\psi^2=\mathbbm{1}$.

Following the above procedure, for each channel choice $(e,c)$, we can 
identify the subset of $\psi$ fields that anticommute with $\sigma$
(dashed lines).
For the Moore-Read state we find,
\begin{equation}
\begin{aligned}
&\begin{minipage}[c]{3.2cm}
\includegraphics[width=\linewidth]{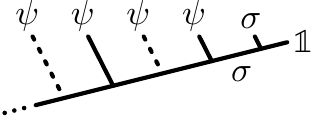}
\end{minipage},&
&\begin{minipage}[c]{3.2cm}
\includegraphics[width=\linewidth]{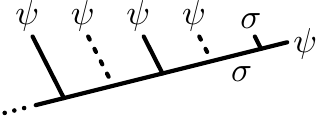}
\end{minipage},\\
&\begin{minipage}[c]{3.2cm}
\includegraphics[width=\linewidth]{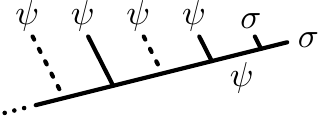}
\end{minipage},&
&\begin{minipage}[c]{3.2cm}
\includegraphics[width=\linewidth]{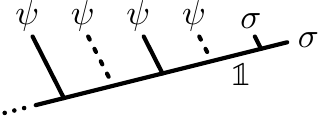}
\end{minipage}.
\end{aligned}
\end{equation}
Counting the dashed lines, we have for 
Eq.~\eqref{eq:successive-shuffling-sign}
\begin{equation}\label{eq:shuffling-mr}
\begin{aligned}
N(\sigma,\mathbbm{1},n)
&=N(\psi,\sigma,n)
=\left\lfloor{\textstyle\frac{n}{2}}\right\rfloor,\\
N(\sigma,\psi,n)
&=N(\mathbbm{1},\sigma,n)
=\left\lfloor{\textstyle\frac{n+1}{2}}\right\rfloor,
\end{aligned}
\end{equation}
where the floor function $\left\lfloor{t}\right\rfloor\in\mathbb{Z}$
denotes the largest integer no greater than $t\in\mathbb{R}$.
Similarly, for the Gaffnian state we have
\begin{equation}
\begin{aligned}
&\begin{minipage}[c]{3.2cm}
\includegraphics[width=\linewidth]{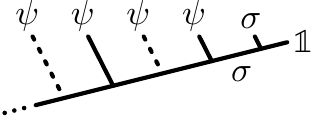}
\end{minipage},&
&\begin{minipage}[c]{3.2cm}
\includegraphics[width=\linewidth]{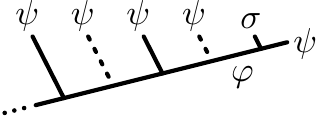}
\end{minipage},\\
&\begin{minipage}[c]{3.2cm}
\includegraphics[width=\linewidth]{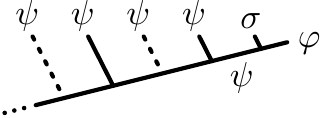}
\end{minipage},&
&\begin{minipage}[c]{3.2cm}
\includegraphics[width=\linewidth]{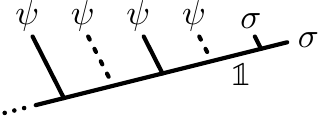}
\end{minipage},\\
&\begin{minipage}[c]{3.2cm}
\includegraphics[width=\linewidth]{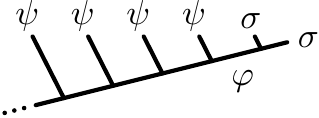}
\end{minipage},&
&\begin{minipage}[c]{3.2cm}
\includegraphics[width=\linewidth]{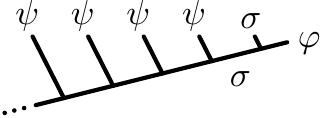}
\end{minipage}.
\end{aligned}
\end{equation}
And the number of anticommuting $\psi$ fields is given by
\begin{equation}
\label{eq:shuffling-gf}
\begin{alignedat}{5}
&N(\sigma,\mathbbm{1},n)
&&=N(\psi,\varphi,n)
&&=\left\lfloor{\textstyle\frac{n}{2}}\right\rfloor,\\
&N(\varphi,\psi,n)
&&=N(\mathbbm{1},\sigma,n)
&&=\left\lfloor{\textstyle\frac{n+1}{2}}\right\rfloor,\\
&N(\varphi,\sigma,n)
&&=N(\sigma,\varphi,n)
&&=0.
\end{alignedat}
\end{equation}
Note that for $(e,c)=(\varphi,\sigma)$ and $(\sigma,\varphi)$, the $\sigma$ 
field commute with all $\psi$ fields without any minus sign.
For the $\mathbb{Z}_3$ Read-Rezayi state we find,
\begin{equation}
\begin{aligned}
&\begin{minipage}[c]{4.0cm}
\includegraphics[width=\linewidth]{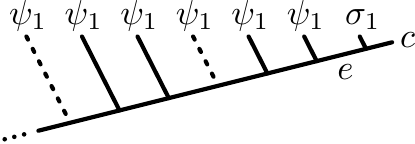}
\end{minipage},\,\,
\bigg[\!\begin{array}{c} c \\ e \end{array}\!\bigg]=
\bigg[\!\begin{array}{c} \mathbbm{1}\\ \sigma_1 \end{array}\!\bigg],
\bigg[\!\begin{array}{c} \sigma_1\\ \psi_1 \end{array}\!\bigg],
\bigg[\!\begin{array}{c} \sigma_2\\ \varepsilon \end{array}\!\bigg];\\
&\begin{minipage}[c]{4.0cm}
\includegraphics[width=\linewidth]{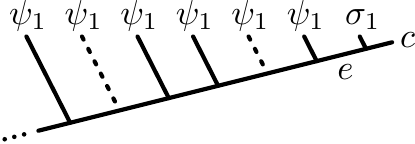}
\end{minipage},\,\,
\bigg[\!\begin{array}{c} c \\ e \end{array}\!\bigg]=
\bigg[\!\begin{array}{c} \psi_1\\ \varepsilon \end{array}\!\bigg],
\bigg[\!\begin{array}{c} \varepsilon\\ \psi_2 \end{array}\!\bigg],
\bigg[\!\begin{array}{c} \sigma_1\\ \sigma_2 \end{array}\!\bigg];\\
&\begin{minipage}[c]{4.0cm}
\includegraphics[width=\linewidth]{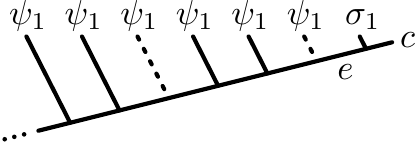}
\end{minipage},\,\,
\bigg[\!\begin{array}{c} c \\ e \end{array}\!\bigg]=
\bigg[\!\begin{array}{c} \psi_2\\ \sigma_2 \end{array}\!\bigg],
\bigg[\!\begin{array}{c} \sigma_2\\ \mathbbm{1} \end{array}\!\bigg],
\bigg[\!\begin{array}{c} \varepsilon\\ \sigma_1 \end{array}\!\bigg].
\end{aligned}
\end{equation}
And the number of anti-commuting $\psi_1$ fields is given by
\begin{equation}\label{eq:shuffling-rr}
\!\!
\begin{alignedat}{3}
&N(\sigma_1,\mathbbm{1},n)=N(\psi_1,\sigma_1,n)=N(\varepsilon,\sigma_2,n)&&
=\left\lfloor{\textstyle\frac{n}{3}}\right\rfloor,\\
&N(\varepsilon,\psi_1,n)=N(\psi_2,\varepsilon,n)=N(\sigma_2,\sigma_1,n)&&
=\left\lfloor{\textstyle\frac{n+1}{3}}\right\rfloor,\\
&N(\sigma_2,\psi_2,n)=N(\mathbbm{1},\sigma_2,n)=N(\sigma_1,\varepsilon,n)&&
=\left\lfloor{\textstyle\frac{n+2}{3}}\right\rfloor.
\end{alignedat}
\end{equation}
Again, we emphasize that the sign structure has contributions from both the 
neutral and the implicit U(1) parts of the direct-product CFT.

\subsection{Putting the pieces together}

We apply the reshuffling formula~\eqref{eq:successive-shuffling-sign} to 
bring the quasihole and the electron fields into time ordering.
A quasihole at position $\chi+i\zeta$ should be placed between the 
orbitals $j_\chi\equiv\left\lfloor{\chi/\gamma}\right\rfloor$ and $j_\chi+1$ 
[Eq.~\eqref{eq:j-chi}].
Starting from the rightmost end of the fusion tree, to reach this time-ordered 
position, the number of electrons it needs to cross is given by the 
number of occupied orbitals with center $x<\chi$.
Consider $m$ fundamental quasiholes with ordering $\chi_1>\cdots>\chi_m$ 
[Eq.~\eqref{eq:multiple-qh}].
The number of \emph{occupied} orbitals with center $x<\chi_l$
is counted by the operator
\begin{equation}\label{eq:electron-counter-k}
\hat{n}_l=\sqrt{\nu}\hat{a}_0+(j_{\chi_l}+1)\nu-\frac{m-l}{k}\nu
\end{equation}
inserted between the $\hat{C}^m$ operators for orbitals $j_\chi$ and 
$j_{\chi}+1$.
The above formula is adapted from the Abelian case [Eq.~\eqref{eq:abelian-qh-sign}].
The extra $k$ in the denominator of the last term reflects the further $k$-fold 
fractionalization of the U(1) charge of a fundamental quasihole in the 
non-Abelian states [Eq.~\eqref{eq:qh-fundamental}].

Now we are finally ready to synthesize the full expression for the non-Abelian 
quasihole operator.
We would like to obtain the MPS description of the conformal block
with $m$ fundamental quasiholes specified by the fusion tree 
\begin{equation}\label{eq:m-qh-tree}
\begin{minipage}[c]{3.5cm}
\includegraphics[width=\linewidth]{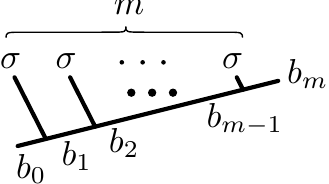}
\end{minipage}.
\end{equation}
Here the quasihole fields are placed at the rightmost end of the linear tree 
structure as in Eq.~\eqref{eq:reordered-tree}, and we have omitted the electron 
branches on the left side.
The topological charges satisfy $b_{l-1}\in \sigma\times b_l$, with
$b_m=\mathbbm{1}$.
In the MPS, the quasiholes are placed in time ordering, and the ${l\text{-th}}$ 
quasihole with fusion context $(b_{l-1},b_l)$ is represented by the insertion of
\begin{equation}\label{eq:B-fundamental-qh}
\hat{U}(\delta\chi_l)\,
\hat{\mathcal{Q}}(b_{l-1},b_l,\hat{n}_l,i\zeta_l)\,
[\hat{U}(\delta\chi_l)]^{-1}\,
(-1)^{N(b_{l-1},b_l,\hat{n}_l)}
\end{equation}
between the $\hat{C}^m$ operators for orbitals $j_{\chi_l}$ and 
${j_{\chi_l}+1}$.
In the above expression, the displacement $\delta\chi_l$ is defined in 
Eq.~\eqref{eq:delta-chi}, 
the projected quasihole operator $\hat{\mathcal{Q}}(e,c,n,i\zeta)$ is 
defined in Eq.~\eqref{eq:qh-after-reshuffling},
the occupied orbital counter $\hat{n}_l$ is defined in 
Eq.~\eqref{eq:electron-counter-k}, and the anticommuting subset 
counter $N(e,c,n)$ defined in Eq.~\eqref{eq:successive-shuffling-sign} has 
values given by Eqs.~\eqref{eq:shuffling-mr}, \eqref{eq:shuffling-gf}, 
and~\eqref{eq:shuffling-rr}.
The above MPS representation preserves the conformal-block normalization for 
the non-Abelian quasiholes, and it is the main result of this paper.

\subsection{Branch cut structure}\label{sec:branch-cut}

\begin{figure}[]
\centering
\includegraphics[]{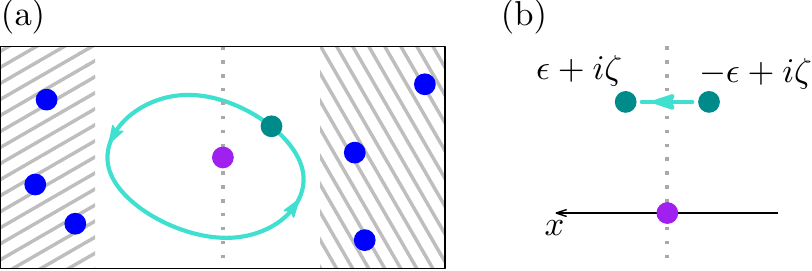}%
\caption{\label{fig:braiding-illustration}
(a) Braiding one quasihole around another. The bystander quasiholes in the 
shaded regions do not participate in the braiding process other than setting 
the fusion channel context. The vertical dotted line marks the location of the 
branch cut discontinuity.
(b) An infinitesimal segment of the braiding path crossing the branch cut from 
above.
}
\end{figure}

As the final note, in this section we examine the monodromy structure of the 
MPS when we braid one non-Abelian quasihole around another 
(Fig.~\ref{fig:braiding-illustration}).
The MPS prescription produces conformal blocks in a particular fusion tree 
basis, where the quasiholes are placed in time ordering at the rightmost end 
of the linear structure before the electrons.
This allows us to focus on the fusion tree segment actively involved in the 
braiding process
\begin{equation}
\begin{minipage}[c]{2.5cm}
\includegraphics[width=\linewidth]{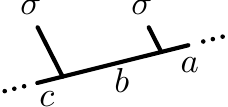}
\end{minipage},
\end{equation}
without worrying about the bystander quasiholes 
(Fig.~\ref{fig:braiding-illustration}) or the electrons, which just define the 
topological charges $(a,c)$.
Due to the time-ordered insertion of the quasiholes,
each conformal block develops a branch cut discontinuity when 
the two quasiholes coincide in the horizontal direction, as illustrated by the 
dotted lines in Fig.~\ref{fig:braiding-illustration}(a).
This discontinuity represents an abrupt \emph{change of basis}
and does \emph{not} correspond to a physical singularity.
To understand how the conformal blocks across the branch cut are related, we 
consider the two points infinitesimally close to the cut, as shown in 
Fig.~\ref{fig:braiding-illustration}(b).
To be specific, we place the stationary quasihole at $\eta_0=0$ and examine the 
infinitesimal segment across the cut above it, from $-\epsilon+i\zeta$ to 
$\epsilon+i\zeta$.
The conformal blocks on the two sides of the cut are given by
\begin{equation}
\begin{minipage}[c]{2.5cm}
\includegraphics[width=\linewidth]{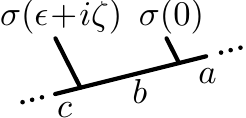}
\end{minipage}
\quad\text{and}\quad
\begin{minipage}[c]{2.5cm}
\includegraphics[width=\linewidth]{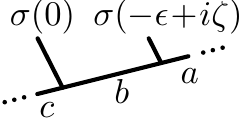}
\end{minipage}.
\end{equation}
In the limit of $\epsilon\rightarrow 0^+$, the conformal blocks on the right side can be 
related via continuity to the twisted trees
\begin{equation}
\begin{minipage}[c]{2.5cm}
\includegraphics[width=\linewidth]{braiding_tree_right.pdf}
\end{minipage}
=
\begin{minipage}[c]{2.5cm}
\includegraphics[width=\linewidth]{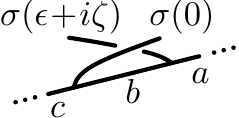}
\end{minipage}.
\end{equation}
Further, we can untwist the fusion trees using the half-braid $B$ matrix of the 
direct-product CFT [Eq.~\eqref{eq:half-braid-familiar}],
\begin{equation}
\begin{minipage}[c]{2.5cm}
\includegraphics[width=\linewidth]{braiding_tree_right_twisted.pdf}
\end{minipage}
=\sum_{d} \big[B^{\sigma\sigma a}_c\big]_{bd}
\begin{minipage}[c]{2.5cm}
\includegraphics[width=\linewidth]{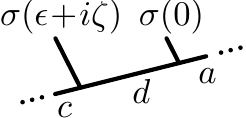}
\end{minipage},
\end{equation}
leading to
\begin{equation}
\begin{minipage}[c]{2.5cm}
\includegraphics[width=\linewidth]{braiding_tree_right.pdf}
\end{minipage}
=\sum_{d} \big[B^{\sigma\sigma a}_c\big]_{bd}
\begin{minipage}[c]{2.5cm}
\includegraphics[width=\linewidth]{braiding_tree_left_d.pdf}
\end{minipage}.
\end{equation}
Similarly, across the branch cut below the stationary quasihole we find
\begin{equation}
\begin{minipage}[c]{2.5cm}
\includegraphics[width=\linewidth]{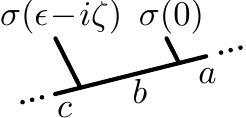}
\end{minipage}
=\sum_{d} \big[B^{\sigma\sigma a}_c\big]_{bd}
\begin{minipage}[c]{2.5cm}
\includegraphics[width=\linewidth]{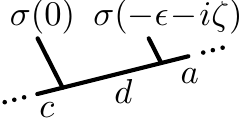}
\end{minipage}.
\end{equation}
Therefore, the MPS conformal blocks across the branch cuts are related by the 
half-braid $B$ matrices of the direct-product CFT.
In other words, the full-braid monodromy of the conformal blocks are 
\emph{concentrated} in the branch cut singularities.
This is a special feature of our fusion tree basis with time ordering,
and it drastically simplifies the microscopic demonstration of the topological 
nature of the quasihole braiding statistics.~\cite{Wu14:Quasihole}
With the branch cuts contributing the CFT monodromy matrix, now we just need 
to show that the braiding process away from the branch cuts accumulates only a 
simple Aharonov-Bohm phase.
For the quasihole wave functions constructed from the conformal correlators,
the latter condition can be further reduced to the exponential convergence of 
the overlap matrix between different conformal blocks at large quasihole 
separations.~\cite{Bonderson11:Plasma}
In an earlier paper, we took advantage of this line of simplifications to 
demonstrate the Fibonacci nature of the $\mathbb{Z}_3$ Read-Rezayi quasiholes.
This exploitation depends crucially on the fact that our MPS construction is a 
literal transcription of the conformal blocks preserving both the monodromy 
structure and the plasma normalization.

\section{Conclusion}

In this paper we have presented a pedagogical derivation of the 
conformal-block MPS for non-Abelian quasiholes.
The procedure is exemplified using the Moore-Read, the Gaffnian, and the 
$\mathbb{Z}_3$ Read-Rezayi states.
Our prescription preserves the monodromy structure and the plasma 
normalization of the conformal blocks, and the resulting MPS explicitly 
manifests the putative quasihole braiding statistics as half-monodromy 
matrices across branch cuts.

We thank P.~Bonderson, V.~Mikhaylov, Y.~Shen, and M.~Zaletel for useful discussions.
BAB, NR, and YLW were supported by NSF CAREER DMR-0952428, NSF-MRSEC DMR-0819860,
ONR-N00014-11-1-0635, MURI-130-6082, Packard 
Foundation, and a Keck grant.
NR was also supported by the Princeton Global Scholarship.
YLW was also supported by LPS-CMTC and JQI-PFC during the final stage of this work.

\appendix

\section{Uniform background charge}
\label{sec:background-charge}

In this Appendix we prove Eq.~\eqref{eq:correlator-cylinder}, namely that the 
inclusion of the uniform background charge [Eq.~\eqref{eq:bc}]
\begin{equation}
\mathcal{O}_\text{bc}=\,\,
:\!\exp\left(-i\frac{\sqrt{\nu}}{2\pi}\int\mathrm{d}^2w\,\,\phi(w)\right)\!:
\end{equation}
in the conformal correlator correctly produces the Landau-gauge Gaussian 
factor with an extra non-holomorphic gauge transform, such that
\begin{equation}
\Psi(z_1,\ldots,z_n)=
e^{i\sum_i x_i y_i}
\Big\langle \mathcal{V}(z_1)\cdots \mathcal{V}(z_n)\,\,
\mathcal{O}_\text{bc}
\Big\rangle
\end{equation}
is a legitimate many-body wave function in the lowest Landau level with the 
Landau gauge. 

The above statement applies to quantum Hall states described by a generic 
direct-product CFT.
Without loss of generality, we can focus on the U(1) part of the electron 
operator, ${\mathcal{V}(z)=\,:\!e^{i\frac{1}{\sqrt{\nu}}\phi(z)}\!:}$,
since the background charge $\mathcal{O}_\text{bc}$ does not couple to the 
neutral CFT.
Further, due to the noninteracting nature of the U(1) boson, we only need to 
consider the contraction of a single electron operator with the background 
charge,
\begin{equation}\label{eq:bc-integral}
\begin{aligned}
e^{f(x,y)}&\equiv
\Big\langle\,\,
:\!e^{i\frac{1}{\sqrt{\nu}}\phi(x+iy)}\!:\,\,\,
:\!e^{-i\frac{\sqrt{\nu}}{2\pi}\int\mathrm{d}x'\mathrm{d}y'\,\phi(x'+iy')}\!:\,\,
\Big\rangle\\
&=\exp\left(
\frac{1}{2\pi}\int\mathrm{d}x'\mathrm{d}y'\,\,
\langle
\phi(x+iy)\phi(x'+iy')
\rangle
\right).
\end{aligned}
\end{equation}

We work on the cylinder geometry with perimeter $L_y$ and inverse radius 
$\gamma=2\pi/L_y$.
Plugging the boson propagator
\begin{equation}
\Big\langle
\phi(z)\phi(0)
\Big\rangle
=-\log\left(\frac{2}{\gamma}\sinh\frac{\gamma z}{2}\right)
\end{equation}
into Eq.~\eqref{eq:bc-integral}, we find
\begin{multline}
f(x,y)=\\
-\frac{1}{2\pi}\int\mathrm{d}x'\mathrm{d}y'
\log\Bigg(
\frac{2}{\gamma}
\sinh\frac{\gamma\,(x+iy-x'-iy')}{2}
\Bigg).
\end{multline}
The integration in the above equation is performed over the cylinder surface 
$(x',y')\in\,{\mathbb{R}\times(-\frac{1}{2}L_y,\frac{1}{2}L_y)}$.
We proceed by separating the real and the imaginary parts of the complex 
logarithm, $\log u=\log|u|+i\text{Im}\log u$.

The real part of the logarithm is nothing but the Coulomb potential on the cylinder 
geometry,~\cite{Aizenman10:Cylinder} in the sense that
\begin{equation}
\nabla^2\log\left|\frac{2}{\gamma}\sinh\frac{\gamma (z-z')}{2}\right|
=2\pi\,\delta(z-z'),
\end{equation}
with cylinder identification $z\sim z+iL_y$. The real part of $f(x,y)$ is thus 
the scalar potential due to the uniform background charge, satisfying
\begin{equation}
\nabla^2\text{Re} f(x,y)=-\int\mathrm{d}x'\mathrm{d}y'\,\delta(x-x')\delta(y-y')=-1.
\end{equation}
Imposing the reflection and the rotational symmetries on the cylinder,
\begin{equation}
\begin{aligned}
\text{Re}f(x,y)&=\text{Re}f(-x,y),&
\partial_y \text{Re}f(x,y)&=0,
\end{aligned}
\end{equation}
we can integrate the above Poisson equation and obtain
\begin{equation}
\text{Re}f(x,y)=-\frac{x^2}{2}.
\end{equation}
As claimed, this correctly reproduces the Gaussian factor in the one-body 
Landau orbital [Eq.~\eqref{eq:Landau-orbital}].

For the imaginary part of the logarithm, we employ the 
identity~\cite{Ogawa09:LogSinh}
\begin{equation}
\text{Im}\log\sinh\frac{\gamma\,(x+iy)}{2}
=\sum_n^{\mathbb{Z}}
\arctan\frac{y+nL_y}{x}.
\end{equation}
It is not hard to see why this holds: starting from
$\text{Im}\log\frac{\gamma}{2}(x+iy)=\arctan\frac{y}{x}$, the periodic 
structure introduced by the hyperbolic sine is matched by the periodic sum 
over $n$.
As a result, the imaginary part of $f(x,y)$ is given by
\begin{equation}
\begin{aligned}
\text{Im}f(x,y)
&=-\frac{1}{2\pi}\int_\text{cyl}\mathrm{d}x'\mathrm{d}y'
\sum_n^{\mathbb{Z}}
\arctan\frac{y-y'+nL_y}{x-x'}\\
&=-\frac{1}{2\pi}\int_\mathbb{R}\mathrm{d}x'\int_\mathbb{R}\mathrm{d}y'
\arctan\frac{y-y'}{x-x'}\\
&=-\frac{y}{2}\int_\mathbb{R}\mathrm{d}x'\,\text{sign}(x-x')=-xy.
\end{aligned}
\end{equation}
Here, we have joined the cylinder integrals in the infinite sum to tile the 
$\mathbb{R}\times\mathbb{R}$ plane, with a symmetric regularization for the 
$\pm\infty$ limits.

To sum up, we have shown that the contraction with the cylinder background charge is 
given by
\begin{multline}
\Big\langle
:\!e^{i\frac{1}{\sqrt{\nu}}\phi(x+iy)}\!:\,\,\,
:\!e^{-i\frac{\sqrt{\nu}}{2\pi}\int\mathrm{d}x'\mathrm{d}y'\,\phi(x'+iy')}\!:
\Big\rangle\\
=e^{-ixy}e^{-x^2/2}.
\end{multline}
This proves that the wave function defined in 
Eq.~\eqref{eq:correlator-cylinder} indeed lives in the Landau gauge with the 
correct Gaussian factor.

\section{Calculating matrix elements}
\label{sec:matrix-element}

In the main text the MPS is described in terms of operators acting on the CFT 
Hilbert space. In this Appendix, we briefly describe the construction of the 
matrix representation of these operators.
A detailed procedure for the calculation of the primary field matrix elements 
has been published in Ref.~\onlinecite{Estienne13:MPSLong}.
Leveraging this result, all we need here is to map the operators of interest 
from the cylinder geometry to the plane.

Under the conformal map $z\rightarrow e^{\gamma z}$ from the 
cylinder to the plane, a primary field $\Phi$ with scaling dimension 
$\Delta_\Phi$ transforms by
\begin{equation}
\hat{\Phi}(z)=\left(\gamma e^{\gamma z}\right)^{\Delta_\Phi}\,
\hat{\Phi}_\text{plane}\!\left(e^{\gamma z}\right).
\end{equation}
Between two generic CFT energy eigenstates 
$|\alpha\rangle$ and $|\beta\rangle$ with $\hat{L}_0$ 
eigenvalues $\Delta_\alpha$ and $\Delta_\beta$,
the matrix element of $\hat{\Phi}(iy)$ reads
\begin{equation}
\begin{aligned}
\langle\alpha|
\hat{\Phi}(iy)
|\beta\rangle
&=\langle\alpha|
e^{iy\gamma\hat{L}_0}
\hat{\Phi}(0)
e^{-iy\gamma\hat{L}_0}
|\beta\rangle\\
&=e^{iy\gamma(\Delta_\alpha-\Delta_\beta)}
\gamma^{\Delta_\Phi}
\langle\alpha|
\hat{\Phi}_\text{plane}(1)
|\beta\rangle.
\end{aligned}
\end{equation}
The planar matrix element 
$\langle\alpha|\hat{\Phi}_\text{plane}(1)|\beta\rangle$
can be further related through algebraic 
manipulations~\cite{Estienne13:MPSLong} to the structure constant associated 
with the $\Phi_\alpha$ terms in the operator product expansion 
$\Phi\times\Phi_\beta$, where $\Phi_\alpha$ and $\Phi_\beta$ are the parent 
primary fields for the states $|\alpha\rangle$ and 
$|\beta\rangle$.
This procedure can be directly applied to the quasihole 
$\hat{\mathcal{Q}}(i\zeta)$.
As for the electron operator, we can reduce the zero mode
$\hat{\mathcal{V}}_0$ [Eq.~\eqref{eq:V-zero-mode}] to a similar form,
\begin{equation}
\begin{aligned}
\langle\alpha|
\hat{\mathcal{V}}_0
|\beta\rangle
&=\gamma^{\Delta_\mathcal{V}}
\int_0^{L_y}\frac{\mathrm{d}y}{L_y}\,
e^{iy\gamma(\Delta_\alpha-\Delta_\beta)}
\langle\alpha|
\hat{\mathcal{V}}_\text{plane}(1)
|\beta\rangle\\
&=\gamma^{\Delta}
\delta_{\Delta_\alpha,\Delta_\beta}
\langle\alpha|
\hat{\mathcal{V}}_\text{plane}(1)
|\beta\rangle.
\end{aligned}
\end{equation}
The relevant structure constants of the neutral CFT for various 
quantum Hall states are documented in the next appendix.

As noted in Sec.~\ref{sec:matrix-product-factorization},
on a cylinder with finite perimeter $L_y=2\pi/\gamma$ the evolution factor 
$e^{-\gamma^2\hat{L}_0}$ allows us to truncate the conformal Hilbert space 
according to $\hat{L}_0$ eigenvalues.
In the actual calculations behind our previous paper,~\cite{Wu14:Quasihole}
we kept the descendant states with $\hat{L}_0<14$ for the Moore-Read and the 
Gaffnian states, and $\hat{L}_0<13$ for the $\mathbb{Z}_3$ Read-Rezayi state.
This is necessary to reach convergence for cylinder perimeter $L_y$ up to 
$25$ magnetic lengths.
The resulting MPS auxiliary space has size up to $3\times 10^4$.
The calculation of physical observables is carried out over the direct product 
of two copies of the auxiliary space,
one for $\langle\!\langle\Psi|$ and one for $|\Psi\rangle\!\rangle$.
The size of this direct-product space can be close to $10^9$.

\section{Operator product expansions}
\label{sec:ope}

In the main text we have listed the field content and the fusion rules for the 
Moore-Read state, the $\mathbb{Z}_3$ Read-Rezayi state, and the Gaffnian wave 
function.
To actually construct the MPS matrices, we also need the full operator product 
expansion with structure constants as noted in the previous appendix.
In the following, we use a shorthand notation
\begin{equation}\label{eq:ope-shorthand}
\Phi_m\times\Phi_n=\sum_l C^l_{mn}\Phi_l
\end{equation}
for the operator product expansion
\begin{equation}
\Phi_m(z)\Phi_n(0)=\sum_l
C^l_{mn}\,z^{h_l-h_m-h_n}
[\Phi_l(0)+O(z)],
\end{equation}
where $h_i$ is the scaling dimension of the primary field $\Phi_i$.
For the Moore-Read state, we have
\begin{equation}
\begin{aligned}
\psi\times\psi&=\mathbbm{1},&
\sigma\times\psi&=\!\frac{1}{\sqrt{2}}\,\sigma,&
\sigma\times\sigma&=\mathbbm{1}+\!\frac{1}{\sqrt{2}}\,\psi.
\end{aligned}
\end{equation}
For the Gaffnian state, we have
\begin{equation}
\begin{aligned}
\psi\times\psi&=\mathbbm{1},\quad
\psi\times\sigma=\frac{1}{\sqrt{2}}\,\varphi,\quad
\psi\times\varphi=\frac{1}{\sqrt{2}}\,\sigma,\\
\sigma\times\sigma&=\mathbbm{1}+C_1\,\varphi,\quad
\sigma\times\varphi=\frac{1}{\sqrt{2}}\psi+C_1\,\sigma,
\end{aligned}
\end{equation}
with the constant $C_1$ given by
\begin{equation}
C_1
=e^{i\pi/4}\left({\textstyle\frac{\sqrt{5}-1}{2}}\right)^{1/4}
\frac{\Gamma(\frac{4}{5})}
{\sqrt{\Gamma(\frac{2}{5})\Gamma(\frac{6}{5})}}.
\end{equation}
Finally, for the $\mathbb{Z}_3$ Read-Rezayi state,
the structure constants are listed in Table~\ref{tab:rr-ope}.

\begin{table}[]
\caption{\label{tab:rr-ope}
The OPE structure constants for the $\mathbb{Z}_3$ Read-Rezayi CFT.
Listed in column $a$, row $b$ is the fusion result $a\times b$ using the 
shorthand notation in Eq.~\eqref{eq:ope-shorthand}.
The constant $C_2$ is given by~\cite{Ardonne07:RR}\\
\begin{minipage}[h]{\linewidth}
\begin{equation*}
C_2=\frac{1}{2}\sqrt{\frac{\Gamma(\frac{1}{5})}{\Gamma(\frac{4}{5})}
\left[\frac{\Gamma(\frac{3}{5})}{\Gamma(\frac{2}{5})}\right]^3}.
\end{equation*}
\end{minipage}
}
\begin{ruledtabular}
\begin{tabular}{c|cc}
$\displaystyle\phantom{\bigg|\!\!}$ & $\psi_1$ & $\sigma_1$ \\\hline\hline
$\psi_1\displaystyle\phantom{\bigg|\!\!}$
& $\frac{2}{\sqrt{3}}\,\psi_2$
& $\sqrt{\frac{2}{3}}\,\varepsilon+\frac{1}{3}\sqrt{\frac{7}{2}}\,\varphi$
\\\hline
$\psi_2\displaystyle\phantom{\bigg|\!\!}$
& $\mathbbm{1}-\frac{\sqrt{26}}{9}\,W$ & $\frac{1}{\sqrt{3}}\,\sigma_2$
\\\hline
$W\displaystyle\phantom{\bigg|\!\!}$
& $\frac{\sqrt{26}}{9}\,\psi_1$ & $\frac{1}{9\sqrt{26}}\,\sigma_1$
\\\hline
$\varepsilon\displaystyle\phantom{\bigg|\!\!}$
& $\sqrt{\frac{2}{3}}\,\sigma_2$ & $\sqrt{\frac{2}{3}}\,\psi_2+\sqrt{C_2}\,\sigma_1$
\\\hline
$\sigma_1\displaystyle\phantom{\bigg|\!\!}$
& $\sqrt{\frac{2}{3}}\,\varepsilon-\frac{1}{3}\sqrt{\frac{7}{2}}\,\varphi$
& $\frac{1}{\sqrt{3}}\,\psi_1+\sqrt{2C_2}\,\sigma_2$
\\\hline
$\sigma_2\displaystyle\phantom{\bigg|\!\!}$
& $\frac{1}{\sqrt{3}}\,\sigma_1$
& $\mathbbm{1}-\frac{1}{9\sqrt{26}}\,W+\sqrt{C_2}\,\varepsilon-\sqrt{\frac{C_2}{21}}\,\varphi$
\\\hline
$\varphi\displaystyle\phantom{\bigg|\!\!}$
& $\frac{1}{3}\sqrt{\frac{7}{2}}\,\sigma_2$
& $-\frac{1}{3}\sqrt{\frac{7}{2}}\,\psi_2+\sqrt{\frac{C_2}{21}}\,\sigma_1$
\end{tabular}
\end{ruledtabular}
\end{table}

\end{document}